\documentclass[10pt, conference, compsocconf]{IEEEtran}
\usepackage{multirow}

\usepackage{caption}

\usepackage{cite}
\usepackage{color}
\ifCLASSINFOpdf
  \usepackage[pdftex]{graphicx}
\else
   \usepackage[dvips]{graphicx}
\fi
\usepackage{amssymb,amsmath}
\usepackage{array}
\begin{document}
%\title{Computational Model for Long Bone Growth Initialization}
%\title{Turing pattern formation on a growing deforming domain as a model for long bone development}
\title{The Control of Branching Morphogenesis}


\author{\IEEEauthorblockN{Dagmar Iber}
\IEEEauthorblockA{Department for Biosystems Science\\ and Engineering (D-BSSE)\\
ETH Zurich\\
Basel, Switzerland\\
E-mail: dagmar.iber@bsse.ethz.ch}
\and
\IEEEauthorblockN{Denis Menshykau}
\IEEEauthorblockA{Department for Biosystems Science\\ and Engineering (D-BSSE)\\
ETH Zurich\\
Basel, Switzerland\\
E-mail: dzianis.menshykau@bsse.ethz.ch}
}


%Telephone: (800) 555--1212, Fax: (888) 555--1212}


\maketitle

\begin{abstract}
Many organs of higher organisms are heavily branched structures and arise by an at first sight similar process of branching morphogenesis. Yet the regulatory components and local interactions that have been identified differ greatly in these organs. It is an open question whether the regulatory processes work according to a common principle and in how far physical and geometric constraints determine the branching process. Here we review the known regulatory factors and physical constraints in lung, kidney, pancreas, prostate, mammary and salivary gland branching morphogenesis, and describe the models that have been formulated to analyse their impacts. 
\end{abstract}


% \fi
\IEEEpeerreviewmaketitle

 \begin{figure}[h!]
\begin{center}
\includegraphics[width=0.61\columnwidth]{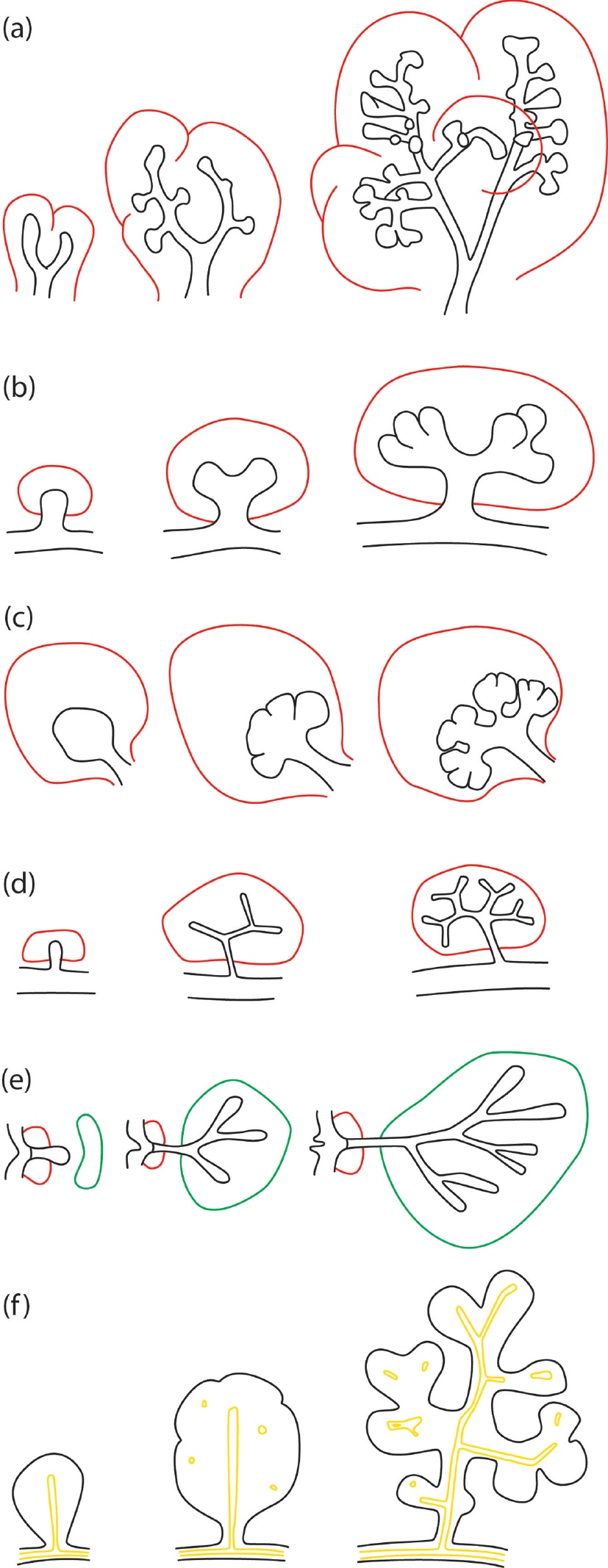}
\caption{Branching Morphogenesis.  Typical branching pattern over developmental time in the (a) lung, (b) ureteric bud, (c) salivary gland, (d) prostate, (e) mammary gland, and (f) the pancreas. The epithelium is shown in black, the mesenchyme in red, the fat pad in the mammary gland in green, and and the lumen in the pancreas in yellow.
 \label{Branching_morphogesis}
 }
 \end{center}
\end{figure}

\begin{figure}[t]
\begin{center}
\includegraphics[width=0.95\columnwidth]{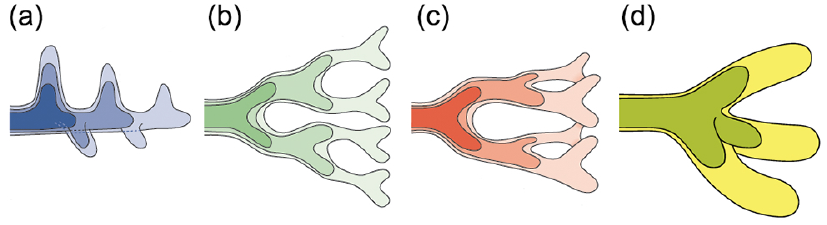}
\caption{Modes of Branching. (a) Lateral branching, (b) planar bifurcation, (c) orthogonal bifurcation, (d) trifurcation. Panels (a)-(c) were reproduced from \cite{Affolter:2009p25219}. \label{Branching_modes}
 }
 \end{center}
\end{figure}

Branching Morphogenesis is observed in many organ systems (figure \ref{Branching_morphogesis}) and in many different species. The branching process in the mammalian lung (figure \ref{Branching_morphogesis}a) and in its analog in flies, the trachea, has been studied in particular great detail. The bronchial tree arises from the sequential use of three geometrically simple modes of branching: domain branching (figure \ref{Branching_modes}a), planar (figure \ref{Branching_modes}b), and orthogonal bifurcation (figure \ref{Branching_modes}c) \cite{Metzger:2008ky}. Trifurcations (figure \ref{Branching_modes}d) have also been documented in the lung \cite{Blanc:2012ea}, but these are much more prevalent in the ureteric bud of the kidney (figure \ref{Branching_morphogesis}b): in the kidney bifurcations and trifurcations dominate, at the expense of lateral branching \cite{Watanabe:2004kr, Meyer:2004de, Costantini:2010p43730}. Similar modes of branching are observed also in glands (figure \ref{Branching_morphogesis}c-f). In this review we will focus on branching during mammalian organogenesis, but ignore neuronal and vasculature branching as well as branching observed in plants. For reviews of these branching systems we refer to \cite{Affolter:2009p25219,OchoaEspinosa:2012ux, Lu:2008p47757, Davies2005, Harrison2010}.

\begin{figure}
\begin{center}
\includegraphics[width=\columnwidth]{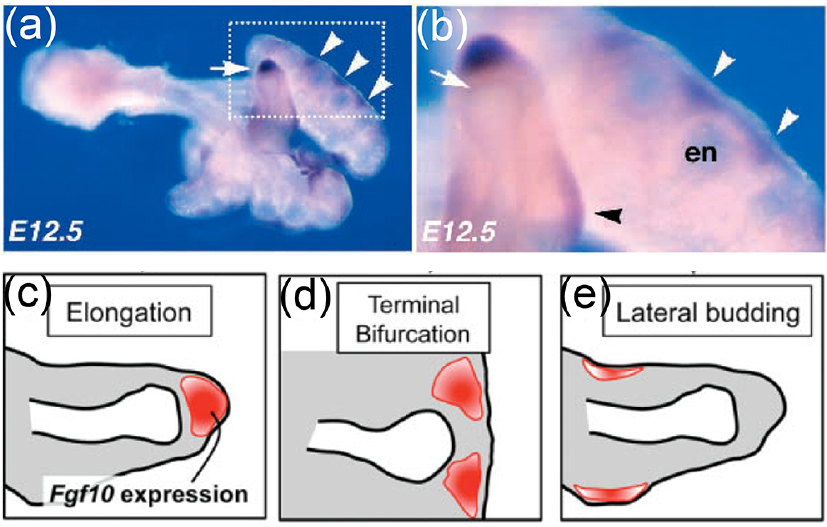}
\caption{FGF10 in the developing lung.  (a,b) \textit{Fgf10} expression at embryonic day (E) 12.5. High expression levels of \textit{Fgf10} are observed in the distal mesenchyme of the tip (white arrow), as well as on the sides of the tips of a bud (white arrowheads). (b) High magnification of the white dotted box in panel a. Note that \textit{Fgf10} expression is absent in the mesenchyme adjacent to the endoderm (en) of the tip. Localized \textit{Fgf10} expression is also observed in the mesenchyme around the stalk (black arrowhead). (c-e) Schematic representations of the spatial distributions of \textit{Fgf10} expression in  panels a,b. \textit{Fgf10} is expressed in the red region. The white area indicates the lumen, the grey area the mesenchyme, the black line in between the epithelium. The outer black lines marks the mesothelium. The three types of spatial distributions of \textit{Fgf10} expression generate different branching modes: (c) elongation, (d) terminal bifurcation, and (e) lateral budding. The figure and the legend were adapted from \cite{Hirashima:2009p43515}; panels a and b were taken from the original publication \cite{Bellusci:1997vq}. 
 \label{Lung_FGF10}
 }
 \end{center}
\end{figure}

Most processes in higher organisms are intricately regulated by a network of signalling factors. To arrive at a branched structure these signalling components would have to form a pattern in space that precedes bud outgrowth as indeed observed in case of fibroblastic growth factor (FGF) 10 in the developing lung bud (figure \ref{Lung_FGF10}a,b). The concentration of FGF10 at the distal tip of the lung bud would then direct elongation (figure \ref{Lung_FGF10}c), while terminal branching would be the result of a split localisation of FGF10 (figure \ref{Lung_FGF10}d), and  lateral branching would result from FGF10 signalling being restricted to spots on the side (figure \ref{Lung_FGF10}e). 

Such patterns can, in principle, emerge from a pre-pattern that goes back to earlier phases of embryonic development or that arises spontaneously from regulatory interactions. A number of theoretical models have been developed to explain how patterns can emerge and how these are read out in space. We will review these concepts in the context of their use in the models for branching morphogenesis. 

\begin{figure*}
\begin{center}
\includegraphics[width=0.7\textwidth]{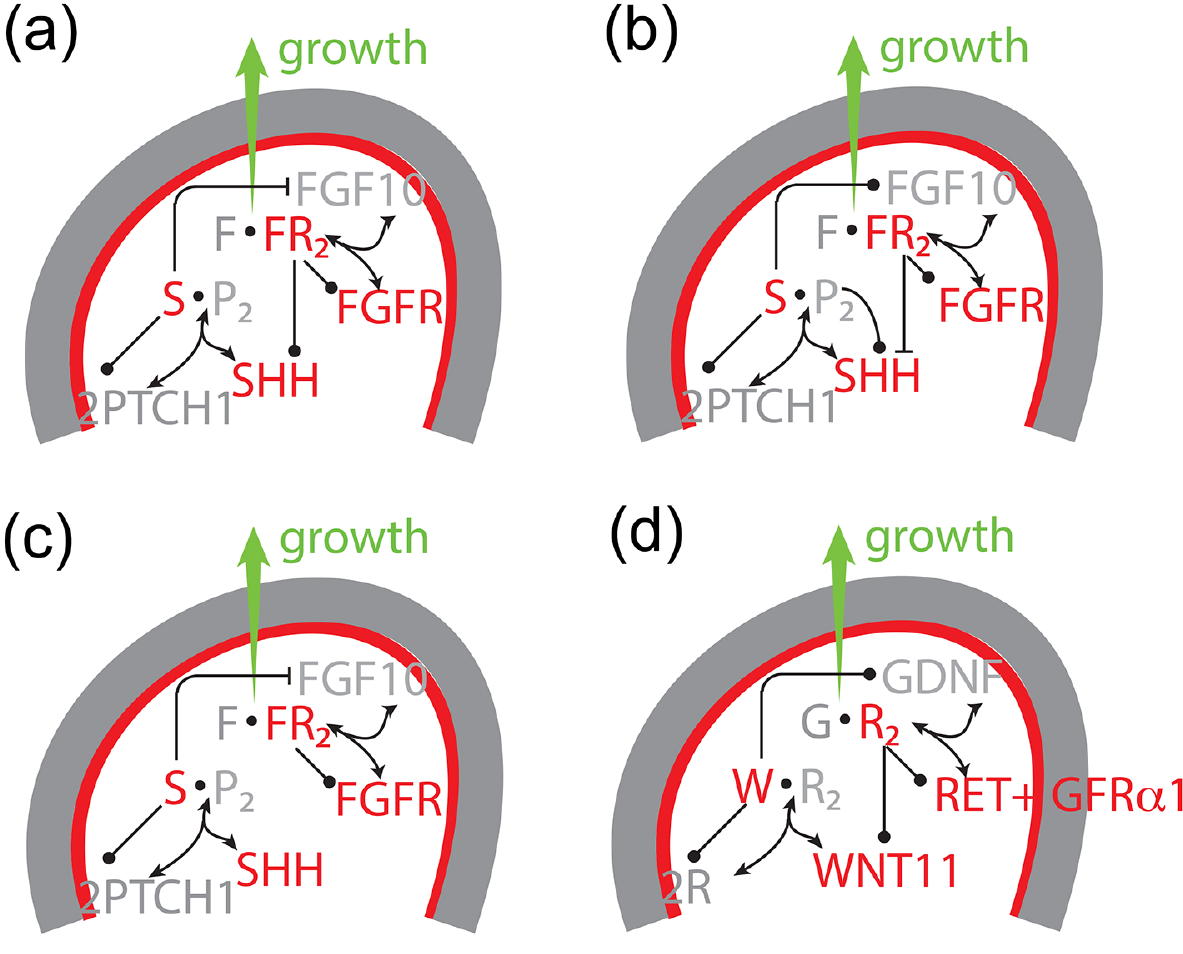}
\caption{Signalling networks in branching morphogenesis. The core signalling networks that have been described to regulate branching morphogenesis in (a) lung and prostate, (b) salivary gland, (c) pancreas, and (d) kidney. In the lung, prostate, salivary gland, and pancreas FGF10 (F) signalling directs outgrowth of the epithelium. \textit{Fgf10} is expressed in the mesenchyme (grey) and binds to its receptor (R) in the epithelium (red). FGF10-bound receptor not only directs outgrowth, but also regulates expression of \textit{Shh} (S) ((a) upregulation in the lung and prostate, (b) downregulation in the salivary gland, (c) no reported regulation in the pancreas). SHH binds its receptor PTCH1 (P) and the SHH-receptor complex, in turn, regulates \textit{Fgf10} expression ((a,c) downregulation in the lung, prostate, and pancreas, (b) upregulation in the salivary gland). All ligand-receptor signalling also upregulates the expression of the receptor. (d) In case of the ureteric bud, GDNF (G) induces bud outgrowth and GDNF-receptor binding stimulates expression of the receptor \textit{Ret} and of \textit{Wnt11} (W) in the epithelium. WNT11, in turn, causes upregulation of \textit{Gdnf} expression in the mesenchyme. 
 \label{Organs_signalling}
 }
 \end{center}
\end{figure*}

While the branching patterns are overall similar, there are important differences (figure \ref{Branching_morphogesis}), and the signalling proteins that control the branching program in the different organ systems are not always the same (figure \ref{Organs_signalling}). Thus FGF10 appears to drive outgrowth of lung buds \cite{Bellusci:1997vq}, prostate \cite{Wilhelm:2006p45512} (both figure \ref{Organs_signalling}a), salivary glands  \cite{Makarenkova:2009ha,Hsu:2010ep} (figure \ref{Organs_signalling}b), and pancreatic buds  \cite{Bhushan:2001ua, Pulkkinen:2003uy} (figure \ref{Organs_signalling}c), while ureteric bud outgrowth is controlled by the TGF-beta family ligand Glial-derived neurotrophic factor (GDNF) (figure \ref{Organs_signalling}d) \cite{Treanor:1996dq, Pichel:1996en, Sanchez:1996cy,Tang:1998vf}, and no single growth factor has yet been defined for the mammary gland \cite{OchoaEspinosa:2012ux}, though FGF receptor 2 is known to be required for ductal elongation in mammary glands \cite{Lu:2008iu}. These signalling proteins are controlled by further proteins and the regulatory networks differ between organs. Given these differences in the signalling networks, also mechanisms based on the interplay of physical forces have been explored.

%These include viscous fingering, where an instability arises at the border of two liquids with different densities, diffusion limited growth, where an instabilities arises during the front movement towards the region of high concentration,  buckling, where an instability arises if the load at the edge of the cylinder is higher than the critical value, and various other mechanisms based on the interplay of signalling and physical forces.

\begin{figure}
\begin{center}
\includegraphics[width=\columnwidth]{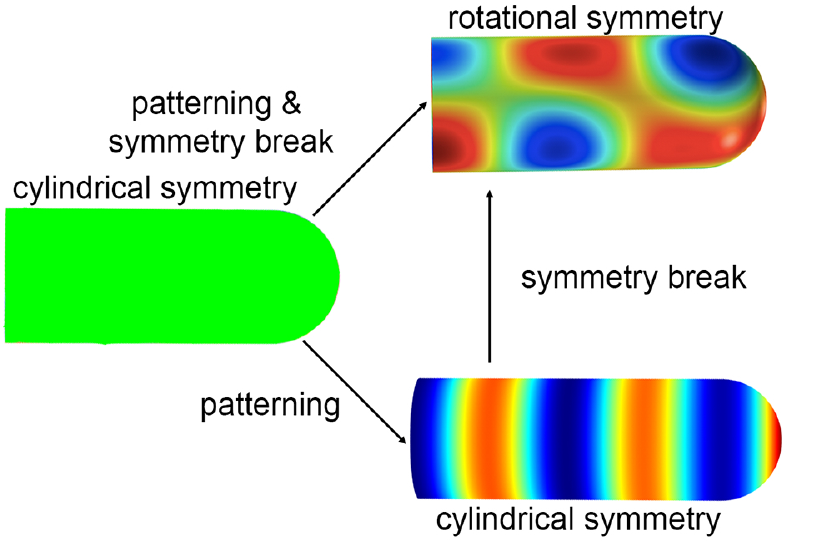}
\caption{Patterning and symmetry break.  A cylinder with a homogenous morphogen concentration exhibits cylindrical symmetry. Patterning mechanisms may introduce stripes or spots. The spotty cylinder exhibits rotational symmetry, and such pattern would support the outgrowth of defined branches.
 \label{fig:mechanisms}
 }
 \end{center}
\end{figure}

Key to all such mechanisms is an instability which results in a patterning and/or a symmetry breaking event. The elongating tube can be regarded as a cylinder with a cap (figure \ref{fig:mechanisms}). A cylinder with a homogenous distribution of morphogens on its surface has a cylindrical symmetry, i.e. if the cylinder is rotated by any angle along the main axis the pattern will not change (figure \ref{fig:mechanisms}). During branching morphogenesis a symmetry breaking event must occur because the outgrowing bud has  approximately cylindrical symmetry while the branching events (i.e bifurcations, trifurcation or lateral branching) change the cylindrical symmetry into a rotational symmetry. Symmetry breaking is a fundamental process and occurs multiple times and at various scales during embryo development as recently reviewed in the collection of Cold Spring Harbor Perspectives in Biology; for the editorial see \cite{Li:2010dv}.

Computational models can help to explore the impact of the signalling interactions, physical forces and domain geometries and thus discern a minimal set of rules and interactions from which the observed pattern can emerge. In the following we will discuss the different models that have been applied to branching morphogenesis of various organs. We will be interested, in particular, in how far similar mechanisms can explain branching morphogenesis in different organ systems. 

%=================================================================================
\section{Lung Branching Morphogenesis}
%=================================================================================

Early lung branching morphogenesis is stereotyped and the lung tree arises from the sequential use of three geometrically simple modes of branching: domain branching, planar and orthogonal bifurcation \cite{Metzger:2008ky}. Transitions from one mode to another are restricted to four routines and lead to three defined sequences which are used to build the entire lung tree \cite{Metzger:2008ky, Affolter:2009p25219}. In the domain branching mode, the lung bud elongates and new buds appear first on one side of the stalk in a direction perpendicular to the main axis of the cylinder and subsequently on another side of the stalk. Planar and orthogonal bifurcations represent two consecutive rounds of bifurcations and differ in the second round of branching, which occurs in the same plane in case of planar bifurcations and orthogonal to the first plane in case of orthogonal bifurcations  \cite{Metzger:2008ky}. Trifurcations (figure \ref{Branching_modes}d) have also been documented recently \cite{Blanc:2012ea}. The domain branching mode is utilized to build the backbone of the respiratory tree, whereas planar bifurcations form the thin edges of the lobes, and orthogonal bifurcations create the lobe surfaces and fill the interior. Only  few variations and errors are observed in wild-type littermates \cite{Metzger:2008ky, Blanc:2012ea}. The regularity of the process implies that the branching process is not random, but tightly controlled by genetically encoded information. It is an open question how a stereotyped branching architecture with millions of branches can be encoded with only a few genes.

\subsection{Fractals}

\begin{figure}[t]
\begin{center}
\includegraphics[width=\columnwidth]{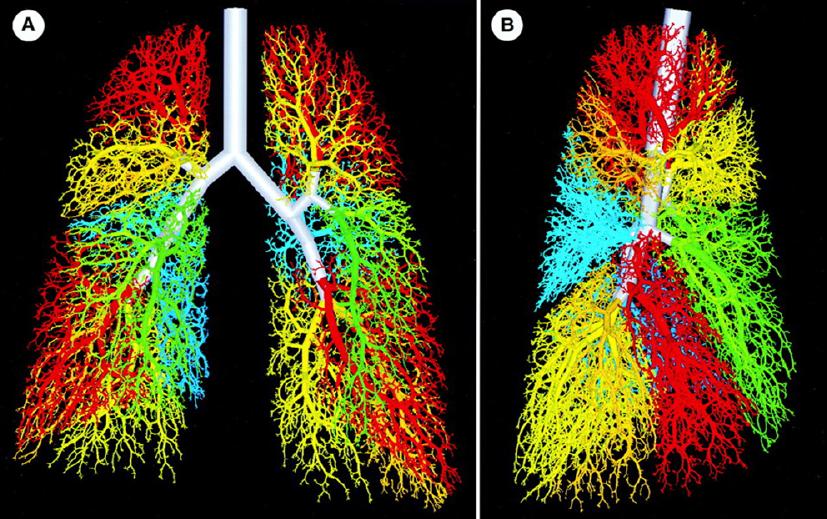}
\caption{A 3-dimensional fractal model of an airway tree with 54,611 branches; branches distal to different segmental bronchi are shown in same color as segmental bronchus. (a): anterior view. (b): right lateral view. The figure and legend were adapted from \cite{Kitaoka:1999uv}. 
 \label{Fig_LungFractal}
 }
 \end{center}
\end{figure}

Fractals are complex structures that can be formed by the repetitive application of a set of simple rules. If the lung was a fractal structure then its complex architecture could result from a set of simple rules as has been demonstrated in many fractal models of the lung. The most sophisticated and accurate of these models was created by Kitaoka and colleagues who required nine basic and four complementary rules to fill the 3-dimensional thoracic cavity with a branched tree that very much resembled that of the lung (figure \ref{Fig_LungFractal}, \cite{Kitaoka:1999uv}). 

To judge whether a geometric figure is indeed a fractal we need to check whether the geometric figures repeat themselves at progressively smaller scales. In case of the lung, dichotomous branching gives rise to two daughter branches with smaller diameter than the mother branch. To obtain a fractal series the shrinkage factor would need to be the same in all generations.

\begin{figure}[t]
\begin{center}
\includegraphics[width=\columnwidth]{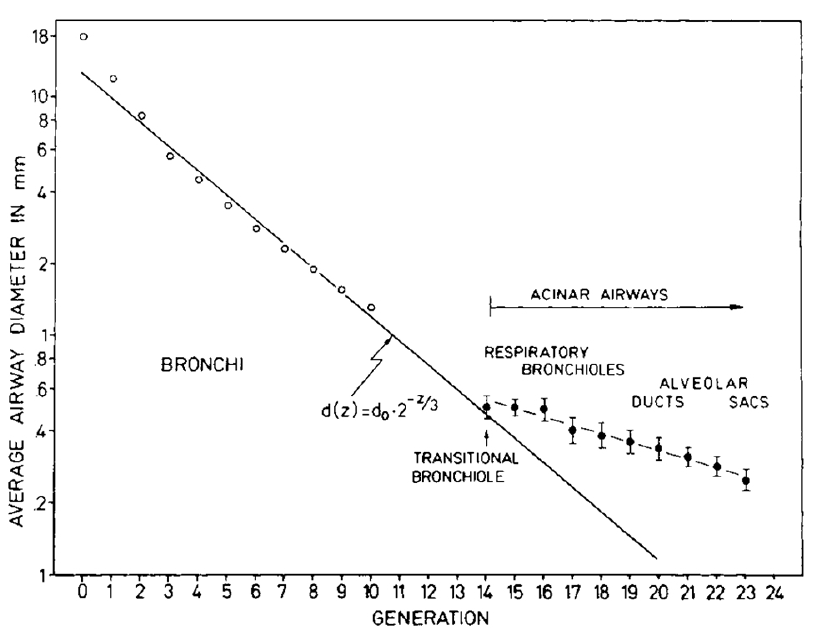}
\caption{A fractal-like organization of the lung.  A logarithmic plot of the airway branch diameter against the branching generation reveals two exponential laws with different scaling factor for conducting and respiratory airways. For the conducting part the relation $d(z) = d(0) \times 2^{-z/3}$ was observed for the diameters $d$ in generation $z$. For such shrinkage factor in the diameter the total volume of all branches remains constant for dichotomous branching, and the entropy generation during breathing is minimal \cite{Wilson:1967wu}. The plot was reproduced from \cite{Bleuer:1988uo}.
 \label{Lung_fractal}
 }
 \end{center}
\end{figure}

Both the length and the diameter of lung branches have been determined for the adult lungs of several mammals \cite{West:1986ty}.  The ratio between diameter and length differs between species, but it is conserved in a given specie (i.e. length / diameter $\sim$ 3 in humans) \cite{Nelson:1990tw}. We can therefore focus on the diameter. Measurements show that the diameter $d_z$ of the lung branches in generation $z$ decrease exponentially with the branching generations (figure \ref{Lung_fractal}). More formally we can write
\begin{equation}\label{Eq_Lung_exponential}
d_z = d_0 \exp{(-\alpha z)} = d_0 q^z, \qquad \alpha =  - \ln q  
\end{equation}
where $q=  \exp{(-\alpha)} <1$ is a constant scaling factor.  We can rewrite this equation as
\begin{equation}
d_z = q d_{z-1}. \nonumber
\end{equation}
The diameter and length in each subsequent generation thus decreases by a constant factor $q < 1$. Such constant scaling law reflects the scale invariance of fractal patterns, i.e. the pattern looks the same no matter on what scale of the structure we zoom in.

In the next step we can determine the fractal dimension of the lung. In the classical interpretation, dimension corresponds to the number of independent coordinates that are required to describe the position of an object in that particular space. To determine a fractal dimension we need to use the so-called Haussdorff dimension. First consider a square. If we scale down each side by a scale factor $r = 1/ 2$, then we require $2^2$ squares to fill the original square. With $r = 1/ n$ we require $n^2$ squares. In 3 dimensions we would require $n^3$ cubes. More generally, if we scale down each side by scale factor $r= 1/ n$ we require $m = \frac{1}{r}^D = n^D$ elements to fill the original space. $D$ is the dimension.
\begin{equation}
m =  \left (\frac{1}{r}\right)^D \quad  \Leftrightarrow \quad D(r) = \frac{\ln{m(r)}}{\ln{1/r}}    \nonumber
\end{equation}

Branching in the lung is dichotomous, and from generation to the next $m =2$  elements thus  emerge. The diameter is scaled by $q= \exp{(-\alpha)}= 2^{-1/3}$ in the first  10-15 generations that generate the conducting airways, i.e. for the airways, which do not participate in gas exchange (figure \ref{Lung_fractal}). Accordingly, the fractal dimension of the lung would be 
\begin{equation}
D_{fractal} = \frac{\ln{m}}{\ln{1/q}} = \frac{\ln{2}}{\ln{2^{1/3}} }= 3, \nonumber
\end{equation}
which would imply that the airway tree fills the entire 3-dimensional thoracic space. Fractal properties in nature have remained controversial because, unlike in mathematical sequences, the geometric figures are observed over a small, finite scale (number of generations) and fitted scaling factors are therefore not particularly reliable \cite{Panico:1995gc}. In fact, a different scaling factor is observed for the lower, respiratory branches (figure \ref{Lung_fractal}, acinar airways), and attempts to simultaneously fit both series have led to a different scaling factor \cite{Mauroy:2004ea}, and to the suggestion of a power-law relation  \cite{West:1986ty}. 

There is, however, independent strong support for the fractal properties of the branching structure of the conducting airways. Thus the required  diameter relation $d_z = 2^{-z/3} d_{0}$ in the conducting part of the airways has been shown to result in an airway architecture that permits breathing with minimal energy wastage (entropy generation) by minimizing the combined effects of dead volume and resistance  \cite{Wilson:1967wu}. Thus the upper part of the lung is dead volume because it must be ventilated without  contributing to gas exchange with the blood. The larger the diameters in this part of the structure, the larger the dead volume, and thus the more energy is wasted in each breath. However, shrinking these diameters results in an increase in air flow resistance. The diameter relation $d_z = 2^{-z/3} d_{0}$ is the one that minimizes the combined contributions of these two effects. The fractal organization of the conducting parts of the bronchial tree thus permits the generation of a space-filling, energy-efficient architecture for ventilation, where all the tips have similar distances from the origin of the airways in the trachea, based on a simple set of rules. While this may explain why such  a bronchial tree evolved, it does not reveal the driving force during development since the embryo does not breath and the developing lung is filled with fluid rather than air.  Moreover, the diameters and lengths in early lung development do not yet exhibit a fractal pattern. So what are the mechanisms that regulate branching morphogenesis?

\subsection{Mechanical Models}

 \begin{figure}[t]
\begin{center}
\includegraphics[width=\columnwidth]{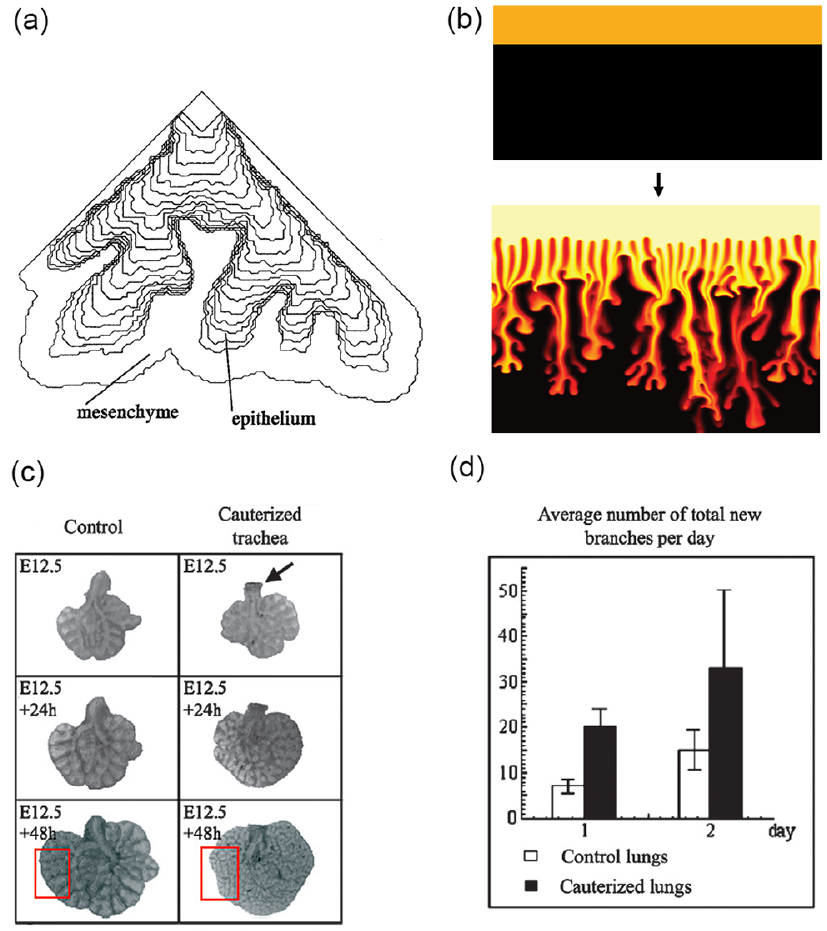}
\caption{Branching as a result of mechanical differences. (a) Branching driven by a viscosity difference between the fluid in the lumen and the mesenchyme; adapted from \cite{Lubkin1995}. (b) Viscous fingering is observed in a system with two immiscible fluids of differing viscosities. The panel was adapted from \cite{Jha2011}.  (c) and (d) The impact of internal pressure on branching; adapted from \cite{Unbekandt2008}. 
 \label{Branching_mechanical}
 }
 \end{center}
\end{figure}

Culture experiments demonstrate that the mesenchymal tissue largely defines the branching pattern when mesenchymal tissue and epithelial tissue of different organs are combined. One striking example came from tissue recombination experiments in which lung mesenchyme induced branching of the ureteric bud with a pattern characteristic of lung epithelium, i.e. with increased lateral branching \cite{Lin:2001vf}. Mesenchymal tissue that either induced or did not induce branching of an epithelial layer in culture was found to differ in its mechanical properties, though the exact differences have not been defined \cite{Nogawa:1987us}. In line with these observations, lung branching was proposed to be driven by the difference in the viscosity of the luminal / amniotic fluid and the mesenchyme, separated by a ''skin" of surface tension, the epithelium \cite{Lubkin1995}. Two fluids with different viscosities exhibit an instability if the more viscous liquid pushes the less viscous one in response to an external force (figure \ref{Branching_mechanical}a); in physics this effect is often referred to as viscous-fingering (figure \ref{Branching_mechanical}b) \cite{Mather1985, Jha2011}. This effect, though physically plausible, has later been recognized as biologically incorrect by the authors \cite{Wan2008, Lubkin:2008iy}, because (1) branching morphogenesis can proceed also without mesenchyme, (2) branching morphogenesis can proceed without growth (hence without growth pressure), and (3) the robustness of the branching process suggests that the branching process is highly controlled, which would not be the case for a near-equilibrium instability. 

In subsequent studies the authors studied a similar model, but interpreted the two (Stokes) fluids as epithelium and mesenchyme, with surface tension in between and focused on the effects of an external ''clefting force", which was assumed to mainly origin from traction forces in the mesenchyme (though epithelial forces were considered as well) \cite{Lubkin2002, Lubkin:2008iy}. The analysis showed that the higher the relative viscosity of epithelium and mesenchyme the wider the resulting clefts and the faster the clefts emerge \cite{Lubkin2002, Lubkin:2008iy}. How the position of the clefting force would be controlled has not been explored, but this could, of course, be regulated by the signalling networks that control branching morphogenesis.

Signalling factors play a key role in branching morphogenesis, and branching of the epithelium in the absence of mesenchyme is observed only if the appropriate signalling factors  (i.e. FGF in case of the lung) are added to the matrigel  \cite{Nogawa:1995wv}. Similarly, while increased internal pressure is a mechanical effect that leads to an increase of lung branching in \emph{in vitro} cultures (figure \ref{Branching_mechanical}c,d), also this effect depends on the activity of FGF signalling, and is not observed in \textit{FgfR2b} null mice, and is reduced in \textit{Fgf10} hypomorphic lung explants \cite{Unbekandt2008}. Signalling networks thus appear to be at the core of the control of branching morphogenesis, and seem to both affect and to be affected by mechanical properties of the tissue. Thus signalling networks regulate the hydraulic pressure during lung organogenesis \cite{Warburton:2010p46351} as well as the actomyosin-mediated contractility of cells, and local differences in the extracellular matrix. Inhibition of the actomyosin-mediated contractility in lung explants decreases branching \cite{Moore2005}, whereas activation of the contractility increases branching \cite{Moore2002}. The impact of the ECM structure on branching morphogenesis has previously been reviewed  \cite{Kim:2012ih}. 

 \begin{figure*}
\begin{center}
\includegraphics[width=0.7\textwidth]{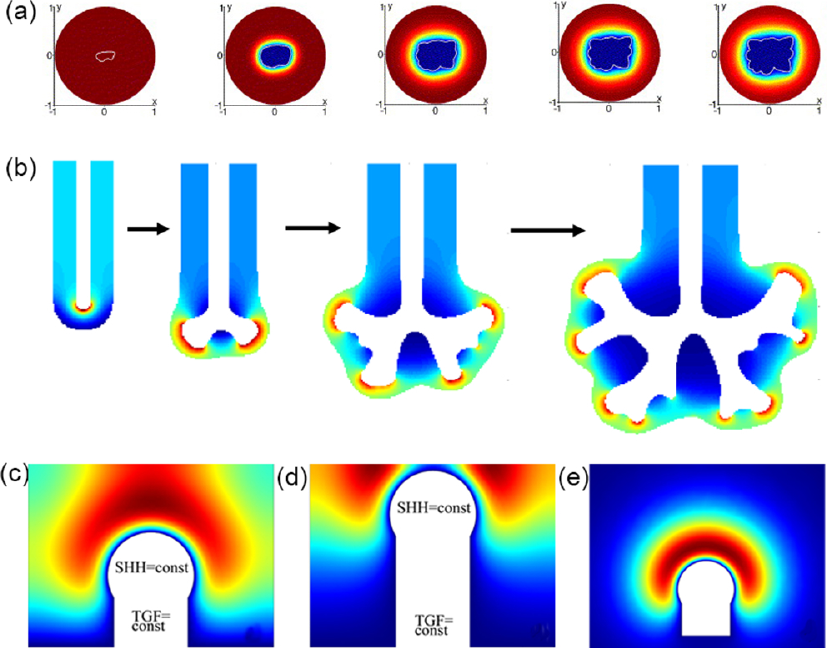}
\caption{Patterning models based on the local variations in the distance of the epithelium from the source of FGF10. 
(a) Branching as a result of diffusion-limited growth. Tissue (delimited by white line) that is placed into a solution with a low concentration of an outgrowth-inducing signalling factor (such as FGF10 in the lung) will degrade the signalling factor and thus induce a concentration gradient (red - high, blue - low). As a result of small irregularities in the tissue shape (white line), some part of the tissue will be closer to the higher concentration and accordingly start to grow out faster, thus experiencing even higher (relative) concentrations. The panel was reproduced from \cite{Hartmann2006}. 
(b) With FGF10 produced mainly in the sub-mesothelial mesenchyme and its receptor produced only in the epithelium, an FGF10 gradient can be expected to emerge. If the FGF10 concentration is homogenous close to the mesothelium and on the epithelium, then the gradient would be steeper the shorter the distance between epithelium and mesothelium. If cells read out gradients rather than concentration, then small differences in the distance could trigger self-avoiding outgrowth of branches. The panel was reproduced from \cite{Clement:2012fj}.  
(c-e) Distance-based mechanism based on SHH and FGF10. SHH is produced only by the epithelium, and  represses \textit{Fgf10} expression at high concentrations. Accordingly,  \textit{Fgf10} expression can be expected to be lower, the closer epithelium and mesothelium \cite{Bellusci:1997vq}.  (c) Assuming that SHH induces \textit{Fgf10} expression at low concentrations, the FGF10 concentration is high as long as  the bud is sufficiently far away from the boundary, thus supporting bud elongation.  (d) As the bud approaches the impermeable boundary, the FGF10 profile splits, thus supporting bifurcating outgrowth.  (e) The bifurcation in panel d requires an impermeable boundary and is not observed on an open domain. The FGF10 distribution in panels (c)-(e) was calculated according to the model presented in \cite{Hirashima:2009p43515}; panels (c)-(e) were adapted from \cite{Menshykau:2012kg}. \label{fig:diffus}
 }
 \end{center}
\end{figure*}

\subsection{Signalling Models}

The core signalling module that controls branching morphogenesis in the lung comprises the two diffusible proteins, FGF10 and SHH (figure \ref{Organs_signalling}a); mutations in both genes and their downstream effectors abrogate branching morphogenesis \cite{Min:1998p48579, Sekine:1999gd, Pepicelli:1998p48620}, while mutations in other genes at most modulate the branching process. FGF10 has been shown to induce the outgrowth of lung buds \cite{Bellusci:1997vq}.  FGF10 and SHH engage in a negative feedback loop, in that FGF10 signalling induces \textit{Shh} expression in the epithelium, while SHH signalling represses \textit{Fgf10} expression in the mesenchyme.  \\

\subsubsection{Diffusion-limited Growth}

The emergence of branches in cultures of mesenchyme-free lung epithelium has been proposed to be the result of diffusion-limited growth  (figure \ref{fig:diffus}a) \cite{Hartmann2006, Hartmann2007, Miura2002}. The diffusion-limited regime of growth is observed when the concentration of the growth factor inducing tissue growth is low and degradation of the growth factor can therefore result in sharp concentration gradients. Small protrusions (which are closer to the source) will then experience higher morphogen concentrations. As a consequence these protrusions will grow faster and thus get even closer to the source. In the experiment, mesenchyme-free lung explants where placed into a matrigel containing FGF. The lung tissue degrades FGF and thus acts as a sink. At sufficiently low FGF concentrations a gradient of FGF emerges with a minimum concentration of FGF in the proximity of the epithelium and the maximum concentration away from the lung explant. For such low concentrations irregular growth of the explant is indeed observed (figure \ref{fig:diffus}a). In the case of high concentrations of FGF in the matrigel no such gradient can emerge and a uniform expansion of the lung explant was observed \cite{Hartmann2006}. When the conditions of diffusion-limited growth were met, both the model and the experiments show that the mechanical strength of the cytoskeleton can only suppress branching. In the chick lung, a branched structure is formed only dorsally while a cyst structure (air sac) is formed ventrally during development. Miura and collaborators suggest that this difference can be accounted to differences in the FGF10 diffusivity that would be sufficiently low only on the dorsal side for branching pattern to emerge by diffusion-limited growth \cite{Miura2009}. 

A further mechanism, similar to the diffusion-limited growth, has been proposed to explain lung branching \emph{in vivo}. Here, however, the authors assume that the FGF concentration on the epithelium is equal everywhere because the epithelium efficiently absorbs the growth factor \cite{Clement:2012fj, Clement:2012hw}. What is assumed to differ is the steady state FGF10 gradient between mesothelium and epithelium. Assuming that FGF10 is uniformly expressed in the sub-mesothelial mesenchyme and FGF receptors are expressed only in the epithelium (such that there is no FGF degradation in the mesenchyme), then the steepness of the steady-state FGF10 gradient between mesothelium and epithelium would depend only on the distance between these tissue layers. If, similar as in the \textit{in vitro} experiments, some part of the bud was closer to the mesothelium than others, then this part of the bud would experience a steeper FGF10 gradient. If cells could sense gradients rather than concentrations then this could, in principle, drive self-avoiding outgrowth of the branch (figure \ref{fig:diffus}b). 
Moreover, given that the distance between source and sink is key to this mechanism, all buds remain at a comparable distance of the mesothelium during growth. This distance is controlled by the mesenchyme proliferation rate relative to the epithelial proliferation rate. The authors also found that the computed branched tree reproduced multiple morphometric characteristics that have been measured in the adult human lung (i.e. the distribution of branch lengths, diameters and the asymmetry ratio) at least qualitatively. However, in experiments, in which FGF10 was added to lung cultures, the mRNA expression of the FGF10 target \textit{Sprouty2} was found to be upregulated \cite{Mailleux:2001vf}. This result is consistent with a model, in which cells respond to the FGF10 concentration, but is difficult to explain with a gradient-based mechanism, as the homogenous addition of FGF10 should reduce the gradient and should thus result in a lower expression level. 

We note that stereotyped branching could be obtained with diffusion-limited growth only, if there was a preceding mechanism in place to generate the same irregularities in the starting geometry for all buds, so that diffusion-limited growth would be initiated always at the same location to obtain the same overall branching pattern.\\

\subsubsection{Distance-based Patterning}

In an alternative model, a different distance-dependent effect has been proposed based on the regulatory interactions between SHH and FGF10 \cite{Bellusci:1997vq}. SHH signalling represses \textit{Fgf10} expression, but \textit{Shh} is expressed only in the epithelium, and \textit{Fgf10} is expressed only in the mesenchyme (figure \ref{Organs_signalling}a). Accordingly, it has been proposed that the inhibitory effect of SHH on \textit{Fgf10} expression will be the stronger, the thinner the mesenchyme. Thus, it has been argued that a locally thinner mesenchyme would lead to the local repression of \textit{Fgf10} expression, and thus to the accumulation of FGF10 on the sides, thereby triggering bifurcating outgrowth. Hirashima and coworkers studied a computational implementation of the model on a static 2-dimensional domain in the shape of lung bud cross-section \cite{Hirashima:2009p43515}. The model  focused on the interactions between FGF10, SHH, and transforming growth factor (TGF)-beta. According to the model TGF-beta is restricted to the stalk and prevents \textit{Fgf10} expression, while SHH is restricted to the tip and enhances \textit{Fgf10} expression at low concentrations and represses \textit{Fgf10} expression at high concentrations. As a result FGF10 is concentrated at the tip as long as the bud is sufficiently far away that the concentration of SHH is low (figure \ref{fig:diffus}c). As the tip grows closer to the impermeable boundary, the local SHH concentration increases and suppresses \textit{Fgf10} expression. As a result, the FGF10 profile splits (figure \ref{fig:diffus}d). While the regulatory interactions in the model are plausible, the mesothelium that surrounds the lung bud is unlikely to present a diffusion barrier. In fact, culture experiments show that protein ligands that are added to the lumen of cultured lungs cannot diffuse to the mesenchyme, but protein ligands that are added to the culture medium can pass through the mesothelium to regulate the mesenchyme \cite{Bragg:2001ws}.  Without an impermeable mesothelial boundary the mechanism does not lead to a bifurcating profile (figure \ref{fig:diffus}e) \cite{Menshykau:2012kg}. We note that this distance-based mechanism would also not explain the lateral branching mode. It is therefore likely that a different mechanism controls branching morphogenesis in the lung. \\

\subsubsection{Diffusion-based Geometry Effect}

\begin{figure}[t]
\begin{center}
\includegraphics[width=0.85\columnwidth]{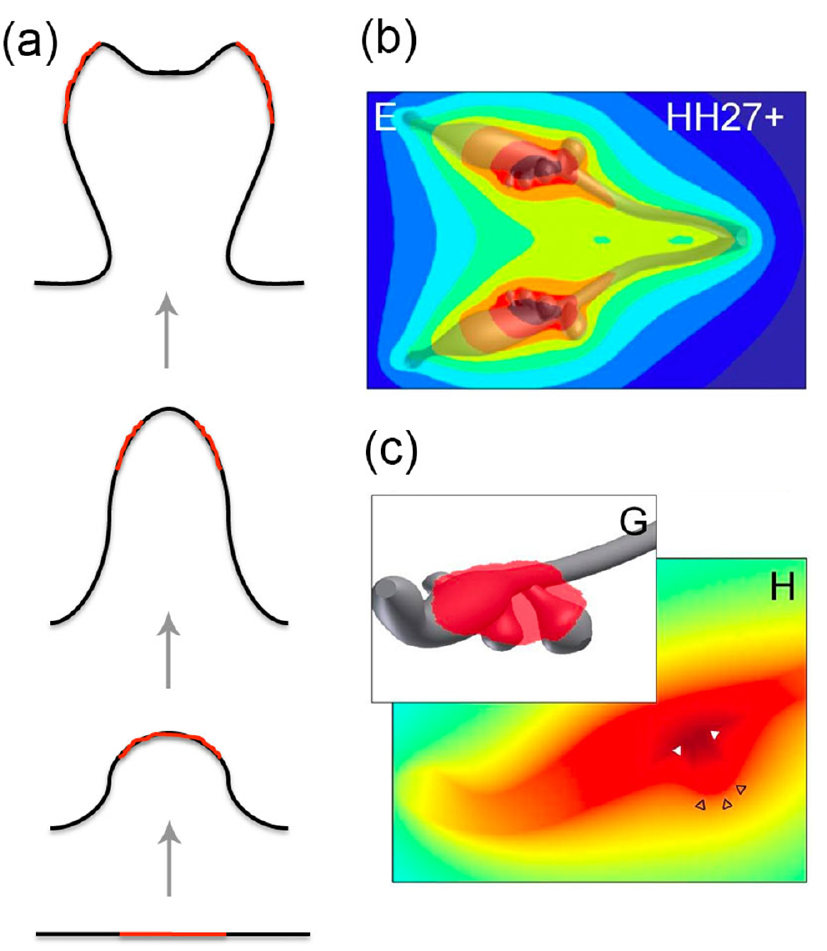}
\caption{Branching as a result of a diffusion-based geometry effect. (a) A cartoon of the proposed geometry-based branching mechanism. As a result of stronger diffusion-based loss at the edges, the concentration of the signal is highest in the center of the domain (red line), and drives the outgrowth of a bud. As the bud elongates, more signal is lost at the tip than at the sides, because of the higher curvature, and a bifurcating concentration profile of the signalling factor emerges. Computational studies confirm that  the geometry effect results in bifurcating concentration profiles, but reveals that it does not support bifurcating outgrowth (Menshykau and Iber, unpublished observation). (b) The simulated concentration profile of a ligand that is uniformly secreted from the epithelium of extracted 3D chicken lung bud (Hamburger-Hamilton stage (HH) 27+) into a large computational bounding box. The concentration profiles were normalized to the highest value (red - highest relative concentration; blue - lowest)
(c) A three-dimensional solid model representation of the region of highest ligand concentration represented by the red shading. Panel (H) shows the morphogen concentration in a cross-section through a bud. The bud stalk and branch point have a local maximum (solid white triangles) whereas the bud tip has a local minimum (empty black triangles) in the predicted ligand concentration. Panels b,c and their legends were adapted from  \cite{Gleghorn:2012el}.    \label{fig:geomtry}}
 \end{center}
\end{figure}

\emph{In vitro} experiments demonstrate that the geometry of the domain can affect its patterning \cite{Nelson:2006gn}. Thus, as a result of diffusion, more signals are lost at the edges of a domain than in its center, if a signal-producing domain is embedded in a non-producing domain, and the signal can diffuse. As a result of this geometry effect, signal then concentrates in the center of the domain (Fig. \ref{fig:geomtry}a). If this factor supports outgrowth of the tissue, then a bud can form. As the bud elongates, more signal is lost at the tip than at the sides, because of the higher curvature, and a bifurcating concentration profile of the signalling factor emerges. Computational studies confirm the emergence of a bifurcating profile as a result of a diffusion-based geometry effect, but also show that it does not support bifurcating outgrowth (Menshykau and Iber, unpublished observation). The geometry effect may nonetheless play an important role in branching morphogenesis by inducing an initial pattern, that can then be ''fixed" by other processes. When 3D shapes of lung bud epithelia were used in a simulation, that had been extracted from early developing chicken lungs, simulated secretion of a ligand from the epithelium into a large computational bounding box (that would model the mesenchyme) resulted in a steady-state concentration pattern that approximately coincided with where the authors would expect branching of secondary bronchi to be inhibited (Fig. \ref{fig:geomtry}b,c) \cite{Gleghorn:2012el}.\\

\subsubsection{Ligand-Receptor-based Turing Mechanism}

\begin{figure}[t]
\begin{center}
\includegraphics[width=\columnwidth]{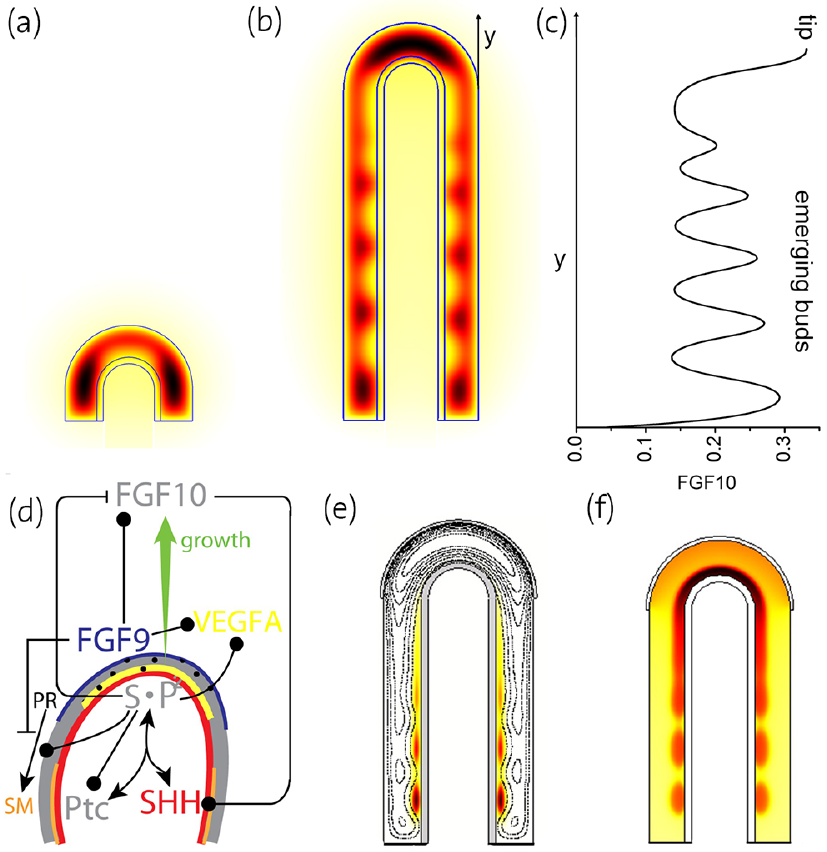}
\caption{A receptor-ligand based Turing model for lung branching morphogenesis.  (a-b) The regulatory network shown in figure 3f gives rise to a Turing pattern and results in  distributions of FGF10 (colorcode, red - highest, black - lowest) and SHH (black $\&$ white contour line plot) on bud-shaped domains as characteristic for (a) bifurcation or (b) lateral branching events. (c) The FGF10 concentration profile along the lung bud. (d) An extended network that includes also FGF9 also reproduces the observed patterns of smooth muscle (SM) formation from progenitors (PR) and \textit{Vegfa} expression during lung branching morphogenesis. (e) Smooth muscles (colour code) emerge in the clefts  between lung buds (contour lines mark FGF10 concentration levels) as the lung bud grows out.  (f) \textit{Vegfa} expression, an inducer of blood vessel formation, emerges in the distal sub-epithelial mesenchyme.  Panels (a-c)  were adapted from  \cite{Menshykau:2012kg} and panels (d-f) from  \cite{Celliere2012}. \label{fig:lung}}
 \end{center}
\end{figure}

We have recently proposed a ligand-receptor based Turing mechanism to explain the stereotyped branching processes in the lung \cite{Menshykau:2012kg}. We showed that the reported  biochemical interactions between FGF10, SHH, and PTCH1 (figure \ref{Organs_signalling}a) give rise to a Turing pattern \cite{MurrayBook} that yields the FGF10 patterns, which represent the two modes of branching in the developing lung: bifurcation and lateral branching (figure \ref{fig:lung}a,b). The initial model analysis was carried out on a 2D slice of the lung bud. Importantly, the Turing pattern not only permitted the emergence of rings (cylindrical symmetry) but also of spotty patterns on a 3D lung bud shape (rotational symmetry) (figure \ref{fig:mechanisms}). Mutations in genes that are not part of the FGF10/SHH core patterning module have been shown to affect the branching pattern. We find that alterations in almost any parameter value can switch the branching mode in our model. The impact of these other gene products on the branching pattern can thus be explained with indirect effects, i.e. by affecting the parameter values of the core model. Another interesting parameter value is the growth rate. We observed bifurcating patterns at low growth speed and lateral branching patterns at fast growth speed. Interestingly, we found that the FGF10 concentration differs in the spots that emerge during lateral branching (figure \ref{fig:lung}c). If the outgrowth speed depended on the FGF10 concentration then these different concentrations might explain the different branching sequences observed for different branches. 

Including FGF9, which is expressed in the distal mesothelium (figure \ref{fig:lung}d, blue line), and which enhances the expression of \textit{Fgf10}, promoted lateral branching over the bifurcation mode of branching as observed in the embryo. Further simulations showed that the expanded core regulatory network was capable of controlling the emergence of smooth muscles in the clefts between growing lung buds (figure \ref{fig:lung}e), and \textit{Vegfa} expression, an inducer of blood vessel formation, in the distal sub-epithelial mesenchyme (figure \ref{fig:lung}f) \cite{Celliere2012}. In how far the vasculature impacts back on the branching pattern is still a matter of investigation. Thus, while ablation of the vasculature impacted on the 3D branching pattern in that branches formed parallel or at a shallow angle instead of perpendicular to the axis, the authors noted that ''inhibition of normal branching resulting from vascular loss could be explained in part by perturbing the unique spatial expression pattern of the key branching mediator \textit{Fgf10} and by misregulated expression of the branching regulators \textit{Shh} and \textit{Sprouty2}'' \cite{Lazarus:2011cp}. 

Another open question concerns the relative width and length of branches. The proposed Turing mechanism would only explain branch point selection and the choice of the branching mode. How the diameters of the developing bronchial tree are determined relative to the length of each branch element is still an open question. Recent experiments suggest that signalling by extracellular signal-regulated kinase 1 (ERK1) and ERK2, a downstream target of FGF signalling, plays an important role because it affects the cell division plane \cite{Tang:2011ii}. A bias in the cell division plane can bias tissue growth in a direction and thus result in either elongation or in a widening of the stalk. Cells that divide parallel to the airway longitudinal axis have lower levels of ERK1/2 signalling and removal of \textit{Sprouty} genes, which encode negative regulators of FGF10 signalling,  results in the random orientation of cell division planes in the stalk and in airways that  are wider and shorter than normal \cite{Tang:2011ii}.

%=================================================================================
\section{Kidney Branching Morphogenesis}
%=================================================================================

Similar to the lung, the kidney collecting ducts form via branching of an epithelial cell layer. During kidney development the ureteric bud invades the metanephric mesenchyme around embryonic day (E) 10.5 \cite{Majumdar:2003tp}. According to culture experiments most branching events in the kidney are terminal bifurcations and to a lesser extent trifurcations, and only $6\%$ of all branching events are lateral branching events \cite{Watanabe:2004kr, Meyer:2004de, Costantini:2010p43730}. FGF10 is not necessary for branching in the lung as branching is still observed in \textit{Fgfr2}-mutant mice, though at a reduced rate \cite{SimsLucas:2009bq}. Core to the branching mechanism appears to be the  TFG-beta family protein GDNF. It  is expressed in the mesenchyme and signals via  its receptor (RET) and co-receptor GDNF family receptor alpha (Gfr$\alpha$)1 in the epithelium (figure \ref{Organs_signalling}d); \textit{Gdnf}, \textit{Ret}, and \textit{Gfr$\alpha$1} null mice do not develop kidneys \cite{Treanor:1996dq,Costantini:2010p43730, Majumdar:2003tp, Pichel:1996en, Pepicelli1997,Sanchez:1996cy, Schuchardt:1994hg}. FGF and GDNF signalling appear to cooperate since activation of FGF10-FGFR2 signalling by knocking out the antagonist \textit{Sprouty} can rescue \textit{Gdnf}$^{-/-}$ and \textit{Ret}$^{-/-}$ mutants, which otherwise fail to develop kidneys \cite{Michos:2010p43732}. Moreover, GDNF and FGF10 signalling have at least in part the same transcriptional targets \cite{Lu2009}. This suggest that the branching mechanisms for lung and kidney are somewhat related in spite of their at first sight different molecular nature. An important difference, however, concerns the core feedback structure. While FGF10/SHH engage in a negative feedback in the lung, GDNF engages in a positive feedback with WNT11 in the ureteric bud \cite{Majumdar:2003tp}.

Beads soaked with GDNF induce the outgrowth of extra ureteric buds in kidney culture explants\cite{Costantini:2010p43730, Majumdar:2003tp, Treanor:1996dq, Pichel:1996en, Pepicelli1997, Sanchez:1996cy}.  Based on the chemoattractive properties of  GDNF \cite{Tang:2002it,Tang:1998vf}, it was suggested that branching of the ureteric bud is caused by the attraction of the tips toward local sources of GDNF \cite{Sariola:2003jo}. Accordingly, in early theoretical work the ureteric bud shape was proposed to be controlled by the interplay of cell proliferation and cell chemotaxis \cite{Hirashima:2009er}. If chemotaxis towards a source of growth factors (i.e. GDNF) dominates relative to the  general growth then the computed branched structure is kinked with clearly discernible buds. On the other hand, if growth dominates relative to chemotaxis then the developing bud is rather round. The model did not attempt to address the question of how a split expression pattern emerges in the first place. Rather, given the split expression of the signalling protein \textit{Gdnf} (as hard-coded in the model), the model addressed the different possible bud shapes.

\begin{figure}[t]
\begin{center}
\includegraphics[width=\columnwidth]{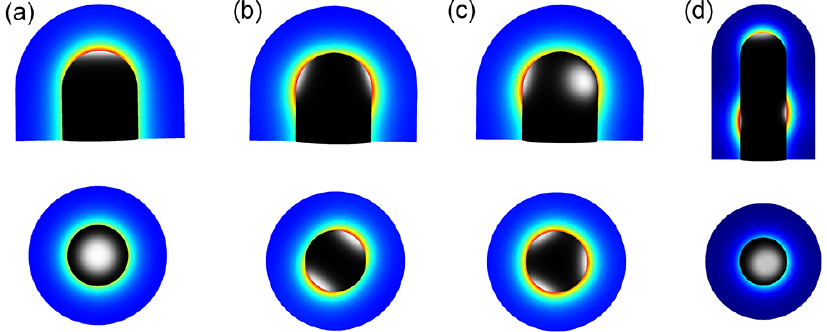}
\caption{Receptor-ligand based Turing model for kidney branching morphogenesis. The reported regulatory interactions shown in figure \ref{Organs_signalling}d can result in self-emerging patterns of the GDNF-bound RET complex in the epithelium (greyscale with white - highest, black - lowest), and \textit{Gdnf} expression in the mesenchyme (rainbow colour code with red - highest, blue - lowest), when solved on a 3D idealized bud-shaped domain. The different patterns can, in principle, (a) support elongation, or support the formation  of  (b) bifurcations,  (c) trifurcations,  or (d) lateral branching. The panels were adapted from \cite{Menshykau:nMxfL07C}. \label{fig:kidney}}
 \end{center}
\end{figure}

\begin{figure*}[t!]
\begin{center}
\includegraphics[width=\textwidth]{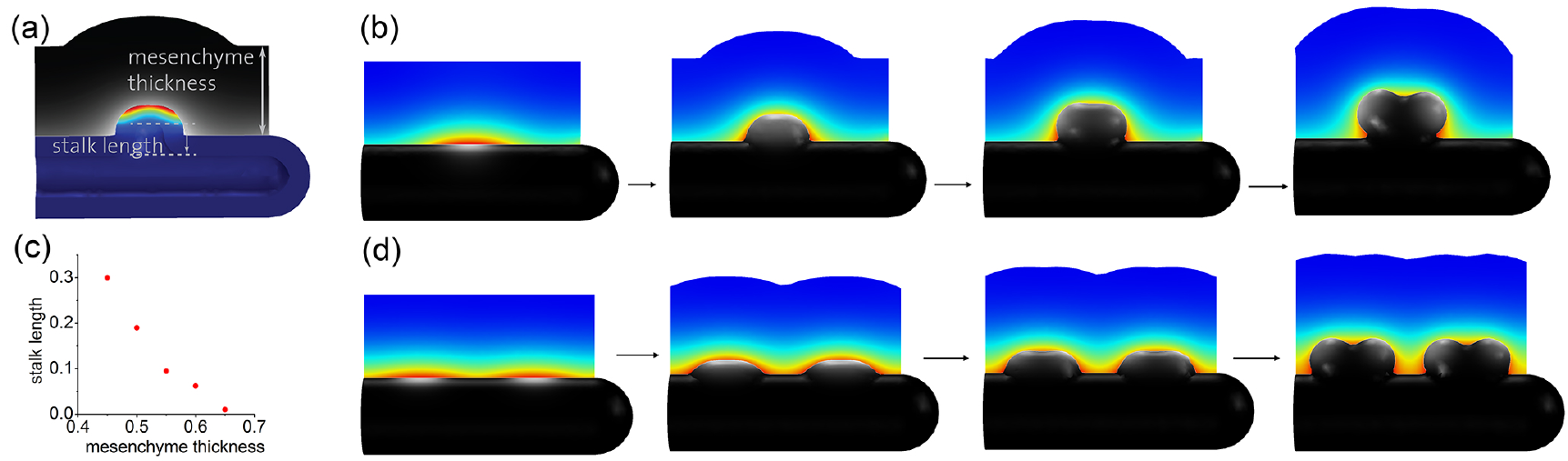} 
\caption{\label{fig:UB} Branching of the Ureteric Bud into the Metanephric Mesenchyme.  (a) The computational domain for the simulation of an ureteric bud after it has started to branch from the Wolffian duct into the metanephric mesenchyme. The double arrows illustrate the stalk length and the mesenchyme thickness. (b) The branching of the ureteric bud into the metanephric mesenchyme in response to GDNF signalling. Here the bud grows out normal to its surface and at a speed proportional to the local concentration of the GDNF-receptor complex, as described in \cite{Iber:2013vf}. (c) The dependency of the length of the stalk to the first branching point on the thickness of the mesenchyme (both as defined in panel a). (d) The branching of the ureteric bud into the metanephric mesenchyme in a Sprouty$^{-/-}$ mutant.  The concentration of  the GDNF-bound RET complex in the epithelium is indicated as greyscale colour code (white - highest, black - lowest); the strength of \textit{Gdnf} expression in the mesenchyme is represented by a rainbow colour code (red - highest, blue -  lowest). The panels were adapted from \cite{Menshykau:nMxfL07C}.}
\end{center}
\end{figure*}

We have recently developed a 3D model for branching morphogenesis in the kidney  \cite{Menshykau:nMxfL07C}. Here we noticed that, similar as in the lung, the reported biochemical interactions between GDNF and its receptors (figure \ref{Organs_signalling}d) result in a Turing pattern. Much as reported for the embryo, split concentration patterns as characteristic for bifurcations (Fig. \ref{fig:kidney}b) and trifurcations (Fig. \ref{fig:kidney}c) dominate in the model for physiological parameter values, while elongation (Fig. \ref{fig:kidney}a) and subsequent lateral branching (Fig. \ref{fig:kidney}d) are rather rare. The patterning mechanism can also support the invasion of the uretric bud into the metanephric mesenchyme (Figure \ref{fig:UB}a,b). It is thus possible that the induction and outgrowth of the ureteric bud from the Wolffian duct works by the same mechanism. Interestingly, in case of thicker mesenchyme the  length of the outgrowing stalk is shorter and branching happens earlier (Figure \ref{fig:UB}c). This is in good agreement with experimental observations in the lung where side-branching was noted to occur when sufficient space becomes available around the circumference of a parent branch  \cite{Blanc2012}.

Turing mechanisms have been proposed for many other biological patterning phenomena, and reproduce the size and geometry-dependence of biological patterns of various complexity \cite{Kondo:2010bx,MurrayBook}. In spite of the great similarity of simulated and real patterns, it remains to be established whether Turing-type mechanisms rather than alternative mechanisms underlie their establishment \cite{Hoefer:tr}. In fact in several cases, Turing-type mechanisms have been wrongly assigned to patterning processes, e.g. to explain the stripy expression pattern of pair-rule genes that  emerge during Drosophila development \cite{Akam:1989kk}. These failures reveal the importance of a careful and comprehensive analysis of the underlying molecular interactions. That this is still not always done is largely due to the lack of sufficient information regarding the chemical properties of the morpho-regulatory proteins (half-life, diffusion rates, endogenous concentrations, activities etc.) and the molecular nature of their interactions (activating, inhibitory etc.). Mutant phenotypes, if considered comprehensively rather than selectively, can also provide valuable information to challenge Turing models.

In the case of lung and kidney branching morphogenesis, we showed that,  for the identified network components, the Turing model reproduces all available mutant phenotypes. In particular, in case of lung branching morphogenesis allelic sequences of the \textit{Fgf10} knock-downs are available. Computational modeling showed that the inter-bud (inter-spot) distance is constant as the \textit{Fgf10} expression rate is reduced from 100$\%$ to 50$\%$, but it greatly increases as the \textit{Fgf10} expression rate decreases below 50$\%$ \cite{Celliere2012}. This is in perfect agreement with the experimental results that show that WT, \textit{Fgf10}$^{\mathrm{LacZ}/+}$, and \textit{Fgf10}$^{+/-}$ have normal phenotypes and \textit{Fgf10} expression between 50-100\%, while \textit{Fgf10} expression in \textit{Fgf10}$^\mathrm{LacZ/-}$ mice is reduced to below 50 \% and an abnormal phenotype with an increased distance between branching points is observed \cite{Mailleux2005, Ramasamy2007}. For kidney branching morphogenesis a \textit{Wnt11}/\textit{Ret} allelic series has been reported \cite{Majumdar2003} and much as in the simulations  the phenotype of \textit{Wnt11} and \textit{Ret} double mutants become more severe as the expression of both \textit{Ret} and \textit{Wnt11} decreases \cite{Menshykau:nMxfL07C}. Finally, in \textit{Sprouty} mutants several ureteric buds branch from the Wolffian duct, as also observed in the simulation (Fig. \ref{fig:UB}d). 

A receptor-ligand based Turing mechanism is thus an attractive candidate mechanism for the control of branching morphogenesis in both lung and kidney, in spite of the different proteins involved. In the following, we will compare the core networks that control branching morphogenesis in glands to the ones in lung and kidney.

%=================================================================================
\section{Branching Morphogenesis in Glands}
%=================================================================================

\subsection{Salivary Glands}
The signalling proteins that regulate branching morphogenesis in the salivary gland are  well characterized, and FGF signalling is necessary for branching morphogenesis (figure \ref{Organs_signalling}b). Thus transgenic mice lacking FGF8, FGF10, FGFR2b, or EGFR do not develop further than the initial bud stage (\textit{Fgf8}-conditional null mice, \textit{Fgf10}-null, and \textit{Fgfr2b}-null mice), or they have substantially fewer terminal buds (\textit{Egfr}-null mice) \cite{Hsu:2010ep}. FGF signalling represses \textit{Wnt} expression \cite{Larsen:2010ki}. WNT signalling upregulates the expression of \textit{Eda} in mesenchymal cells and EDA upregulates \textit{Shh} expression in epithelial cells via NF-kB signalling \cite{Haara:2011kn}. SHH signalling upregulates \textit{Fgf8} expression  \cite{Jaskoll:2004eo}, which in turn upregulates \textit{Fgf10} and \textit{Shh} expression \cite{Jaskoll:2004hl}. Removal of the \textit{Eda receptor} or \textit{Shh} results in dysplasia \cite{Jaskoll:2004eo}. Knock-out of \textit{Bmp7} results in reduced branching, and BMP7 is able to rescue  branching in salivary glands treated with the FGFR signalling inhibitor SU5402, suggesting that BMP7 may be downstream of FGFR signalling, or in a parallel pathway \cite{Patel:2006ed}. The receptor-ligand interactions are thus very similar to those in the lung (e.g. FGF10-FGFR2b or SHH-PTCH1) and kidney (i.e. WNT and FGF10), and mesenchyme from lung buds can induce branching of cultured submandibular epithelium \cite{Nogawa:1987us}. Given these parallels, a similar Turing-based mechanism may control branching also in the salivary gland. Experiments, in which the FGF10 diffusivity was modulated by alteration of the binding affinity of FGF10 for heparan sulfate, highlight the relevance of a diffusion-based mechanism as they reveal an effect on the branching pattern (elongation versus branching) in the salivary gland \cite{Makarenkova:2009ha}.

\subsection{Pancreas}

In contrast to the classical epithelial budding and tube extension observed in other organs, a pancreatic branch takes shape as a multi-lumen tubular plexus and coordinately extends and remodels into a ramifying, single-lumen ductal system \cite{Villasenor:2010iz}. This initial tube formation takes place  from E9.5 to E12.5. During a secondary transition from E13 to birth, branching morphogenesis is observed \cite{Benitez:2012hc}. \emph{In vivo} time lapse imaging reveals that the main mode of branching  during pancreas branching morphogenesis is lateral branching (86$\%$); terminal bifurcations are also observed, however less often ($14\%$) \cite{Puri:2007jp}. The lateral branching mode deployed during pancreas development is simpler than that in the case of lung branching - new buds appear predominantly on one side of the outgrowing tip. However, this might be an artifact of the culture experiment, e.g. lung branching patterns are known to be different \emph{in vivo} and \emph{in vitro}. Similar to lung branching morphogenesis both FGF10/FGFR2b \cite{Bhushan:2001ua, Pulkkinen:2003uy}, and SHH/PTCH1 receptor-ligand modules play an important role during pancreas branching morphogenesis, with Hedgehog signalling repressing \textit{Fgf10} expression (figure \ref{Organs_signalling}c) \cite{Kawahira:2003p43684}. Two factors, BMP4 and FGF10, promote pancreatic morphogenesis at the primary stages of the organogenesis. \textit{Bmp4} is the first to be expressed and promotes budding formation, whereas \textit{Fgf10} is expressed later and promotes (among other things) branching of the pancreatic bud \cite{Bhushan:2001ua}. Ephrin signalling has also been implicated in pancreatic branching as removal of the \textit{EphB2} and \textit{EphB3 receptors} results in shortened branches and smaller pancreata  \cite{Villasenor:2010iz}. The developing pancreas displays predictable trends in overall shape and in the elaboration of specific branches \cite{Villasenor:2010iz}. The early stages have been simulated using a fully executable, interactive, visual model for 4D simulation of organogenic development \cite{Setty:2008ea}. A computational model of the later developmental stages, including the branching morphogenesis of the exocrine pancreas, is still required as is a model of the processes that result in the differentiation of the pancreatic tissue into ducts and enzyme secreting acinar cells at the tips of the branches.

\subsection{Prostate}
Circulating androgens initiate the development of the prostate from the urogenital sinus \cite{Wilhelm:2006p45512}. The androgen receptor is expressed in the mesenchyme and is necessary for budding of the urethral epithelium \cite{Wilhelm:2006p45512}. Androgens have been shown to induce the expression of \textit{Fgf10} and FGF receptor 2 (\textit{Fgfr2})-IIIb in the urethrea \cite{Petiot:2005dr}. FGF7 and FGF10, expressed in the mesenchyme, bind to FGFR2 on epithelial cells, which leads to the induction and maintenance of \textit{Shh} expression  \cite{Wilhelm:2006p45512}. SHH signalling, in turn, downregulates \textit{Fgf} expression \cite{Wilhelm:2006p45512}. Moreover, ductal branching and budding are inhibited by the mesenchymal signalling factors BMP4 and BMP7, and stimulated by the antagonist of these BMPs. The regulatory loop is thus very similar to the one in the lung (figure \ref{Organs_signalling}a), and FGF10 has been found to be essential for the development of the fetal prostate \cite{Donjacour:2003ty}. It should be noted that most of the ducts remain unbranched until birth in rodents but, subsequently, epithelial-mesenchymal interactions result in further elongation and branching morphogenesis.

\subsection{Mammary Glands}

The development of the mammary proceeds in three stages: a rudimentary gland develops in the embryo and remains quiescent until puberty \cite{Macias:2012iv}. During puberty extensive branching occurs, and the mammary glands will undergo further rounds of branching during pregnancy. In male mice, the mesenchyme surrounding the stalk continues to condense until it severs the bud resulting in a greatly diminished ductal system \cite{Kratochwil:1976tq}. Signalling of parathyroid hormone-like hormone (PTHLH) via Type1 PTH/PTHLH receptor (PTH1R) appears to play a key role in the initial process of generating a rudimentary ductal system before birth \cite{Wysolmerski:1998vu}. \textit{Pthlh} is expressed in the epithelium and signals through its mesenchymal receptor PTH1R to modulate WNT signalling, and to induce the BMP receptor-1A (BMPIRA) in the mesenchyme \cite{Hens:2007cv}. BMP4 signalling can rescue ductal outgrowth in cultures of  \textit{Pthlh}$^{-/-}$ mammary buds \cite{Hens:2007cv}. 
 
Similar to the lung and the kidney, epithelial-mesenchymal signalling by FGFs, EGFs and WNTs plays an important role in controlling branching morphogenesis \cite{Macias:2012iv, Gjorevski:2011hg}. FGF receptor 2 is required on epithelial cells for ductal elongation \cite{Lu:2008iu}, but a unique chemoattractant like FGF10 in the lung and salivary gland \cite{Park:1998fz} or GDNF in the kidney \cite{Tang:1998vf, Tang:2002it} has not been identified \cite{OchoaEspinosa:2012ux}. TGF-$\beta$ signalling, on the other hand, has been shown to be important to reduce branching and thereby restrict branch formation \cite{Pierce:1993va, Nelson:2006gn}. TGF-$\beta$ signalling has been shown to induce the deposition of extracellular matrix \cite{Pierce:1993va, Nelson:2006gn} and to affect basal cell proliferation via roundabout 1 (ROBO1), SLIT2, and WNT signalling  \cite{Macias:2011jt}; the non-canonical WNT signalling member WNT5a is necessary for the effects of TGF-$\beta$ on branching morphogenesis \cite{Roarty:2007hq}. 

Culture experiments support a similar control mechanism for branching morphogenesis as in the other organs, as mesenchyme from the salivary gland  induces branching of the mammary gland epithelium, even though the branched structure and lobe resemble then that of the salivary gland \cite{Kratochwil:1969tj}. Contrary to the lung and the kidney, branching in the mammary glands is, however, not stereotyped \cite{OchoaEspinosa:2012ux, Gjorevski:2011hg}. The mechanisms that have been proposed to control branching morphogenesis in the mammary gland, have so far mainly focused on mechanical constraints, as provided by interactions with the extracellular matrix \cite{Kim:2012ih, Pozzi:2011eu}, and on diffusion-based geometry effects  (Fig. \ref{fig:geomtry}a) \cite{Nelson:2006gn, Nelson:2012ku}. Experiments support an influence of mechanical stress on branching morphogenesis in mammary glands \cite{Gjorevski:2010kr}, but the composition of the ECM is likely the result of regulation by signalling networks  \cite{Pierce:1993va, Nelson:2006gn}. \\

 %=================================================================================
\section{A general mechanism for the control of stereotyped branching?}
%=================================================================================

The early branching events differ between the organs (figure \ref{Branching_morphogesis}), but are highly stereotyped (at least in lung and kidney), and must therefore be carefully controlled. A number of mechanisms have been proposed, but most fail to meet all key aspects of such a branching mechanism, which are: 1) the production of stereotyped pattern from noisy initial conditions, 2) pattern stability (or pattern fixation) during outgrowth, such that the pattern can support the outgrowth of the branch, and 3) the ability to control branching morphogenesis in different organ systems with different signalling networks. The receptor-ligand based Turing mechanism meets all these conditions  \cite{Celliere2012,Menshykau:2012kg,Menshykau:nMxfL07C}. In particular, it  works with a range of signalling networks, because all signaling involves the interaction of secreted ligands with some receptors (figure \ref{Organs_signalling}), and any ligand-receptor pair that fulfills the following three conditions can give rise to Turing-type reaction kinetics:

\begin{enumerate}
\item The ligand must diffuse faster than its receptor, which is generally the case for soluble ligands and membrane receptors  \cite{Choquet:2003fk,Ries:2009p20248,Kumar2010, Hebert2005}.
\item Receptor and ligand must interact cooperatively, which is typically the case when multimeric components bind each other.
\item Receptor-ligand binding must result in increased receptor production. This has been documented in case of SHH/PTCH1  \cite{Weaver2010}, and GDNF/RET \cite{Pepicelli1997, Lu2009, Costantini2010}. In case of FGF10 both up- and down-regulations have been reported  \cite{Bansal1997, Estival2010, Ota2010, Zakrzewska:2013gt}. BMPs as well as other ligand-receptor systems would also meet the requirements \cite{Merino:1998ha, Badugu:2012ho}. 
\end{enumerate}

    \begin{figure}[t]
\begin{center}
\includegraphics[width=\columnwidth]{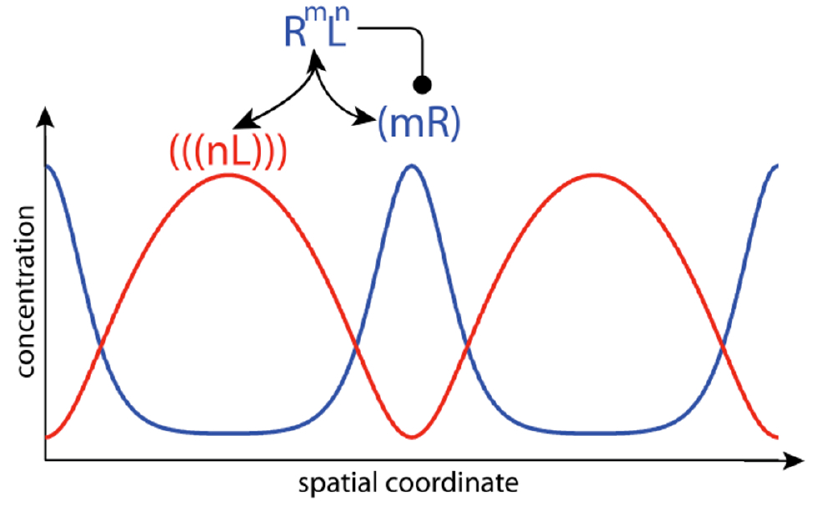}
\caption{Cooperative receptor-ligand interactions can give rise to Turing patterns. The depicted receptor-ligand interaction can result in spatial patterns via Schnakenberg-type reaction kinetics. Here $m$ receptors ($R$) and $n$ ligand molecules ($L$) (with $m+n >2$) need to bind to form the complex $R^mL^n$. The receptor ligand complex then up-regulates the receptor concentration (by increasing its expression, by limiting its turn-over, or similar). To obtain Turing patterns, ligands must diffuse much faster than their receptors. As characteristic for Schnakenberg-type Turing pattern, the highest receptor and ligand concentrations are then observed in different places.  \label{Receptor_ligand_Turing}
 }
 \end{center}
\end{figure}

The ligand-receptor based Turing mechanism is different from the classical activator-inhibitor Turing mechanism, also known as Gierer-Meinhardt Turing mechanism \cite{Gierer:1972vq}, which is the most commonly studied Turing mechanism.  Unlike in the activator-inhibitor mechanism, only one ligand is required in the here-proposed mechanism, and the two components (ligand and receptor) assume their highest (and lowest) concentrations in different places (figure \ref{Receptor_ligand_Turing}). The ligand-receptor mechanism is based on Schnakenberg-type reaction kinetics \cite{Schnakenberg:1979td}, which correspond to the activator-depleted substrate Turing mechanism  \cite{Gierer:1972vq}. 

Turing mechanisms have been proposed for many systems, and also an alternative Turing-based model has previously been proposed that achieved branching morphogenesis by a combination of patterning and fixation events via cell differentiation \cite{Meinhardt:1976vc}. Further experimental studies are now required to carefully test the proposed  mechanisms.  Turing patterns, much as the other proposed mechanisms, are very sensitive to the geometry of the domain. Simulations of the models on realistic embryonic geometries would therefore present a good test to the model. 
Developmental sequences of 3D geometries can now be obtained with optical projection tomography (OPT) \cite{Sharpe2002}, two photon or light sheet microscopy \cite{Verveer:2007ck, Truong:2011cn}. Software packages are readily available to construct the 4D geometric series and to efficiently simulate the signalling models on the growing and deforming domains \cite{Menshykau:2012vg,Germann:bT_kMV7D, Iber:2013vf}. The correct mechanism should be able to predict the branch points. 

Current models typically consider tissue as a continuum, without cellular resolution. Cell boundaries, however, can have important effects, in particular in the ligand-receptor based Turing mechanism, where receptor is restricted to the surface of single cells. While we have  previously shown that similar patterns can be obtained also on a static, cellularized domains \cite{Menshykau:2012kg}, it will be important to develop simulation tools that permit the efficient simulation of patterning in 3D cellular systems. Here it will be important to include cell growth, diffusion of soluble factors as well as diffusion restrictions imposed by cell boundaries to membrane and cytoplasmic proteins.

Finally, it will be important to address the issue that classical Turing models produce patterns only for a tiny parameter space. For such small parameter space it is difficult to explain how patterns could emerge in the first place and how they could be preserved during evolution. \\

\section{Conclusion}

Branching morphogenesis has long fascinated biologists and theoreticians and many different effects have been defined that all impact on the branching process. We propose that Turing patterns as a result of  receptor-ligand interactions constitute a general core mechanism that controls branching morphogenesis in the different organs, as well as other patterning processes in the various developmental systems. Much as complex fractal patterns can be generated with few rules, Turing mechanism can, in principle, implement a complex stereotyped sequence of branching events based on few interacting proteins. It should be noted that fractals and Turing patterns are otherwise very different, and Turing mechanisms do not give rise to fractal patterns. An additional mechanism must therefore define the diameter of the tubes to eventually give rise to a fractal-like sequence in the adult. \\

Many of the other discussed effects are likely to affect the patterning process without being its key controller. One such important effect is likely the diffusion-based geometry effect (figure \ref{fig:geomtry}) \cite{Nelson:2006gn}, which biases the pattern, but which cannot support bifurcating outgrowth by itself. Moreover, the distance between the \textit{Fgf10}-expressing distal mesenchyme (close to the mesothelium) and the \textit{Shh}-expressing epithelium can be expected to affect the exact lung branching pattern without being responsible for the bifurcating concentration profiles \textit{per se} \cite{Bellusci:1997vq}. Similarly, mechanical differences are clearly affecting the branching process.  How these arise and how they are controlled in space and time will be an important direction of future research. \\

A key open question concerns the mechanism that mediates branch outgrowth once  an initial symmetry break in the cellular signalling has defined the branch points. This is likely the  result of a combination of local changes in the proliferation of cells (in particular their cell division plane), cell deformation, and cell migration \cite{Tang:2011ii, Kim:2013db, Schnatwinkel:2013kia, Hsu:2013hu}.  \\

\section*{Acknowledgment}
%=================================================================================

The authors acknowledge funding from an SNF Sinergia grant, the SNF SystemsX RTD NeuroStemX, a SystemsX iPhD grant, and an ETH Zurich postdoctoral fellowship to D.M.. The authors are grateful to Philipp German for drawing Figures 1 and \ref{Branching_modes}d, and to Markus Affolter, Odysse Michos, and Jannik Vollmer for critical reading of the manuscript.

\bibliographystyle{./IEEEtranBST/IEEEtran}

%\bibliography{./papers}

\begin{thebibliography}{100}
\providecommand{\url}[1]{#1}
\csname url@samestyle\endcsname
\providecommand{\newblock}{\relax}
\providecommand{\bibinfo}[2]{#2}
\providecommand{\BIBentrySTDinterwordspacing}{\spaceskip=0pt\relax}
\providecommand{\BIBentryALTinterwordstretchfactor}{4}
\providecommand{\BIBentryALTinterwordspacing}{\spaceskip=\fontdimen2\font plus
\BIBentryALTinterwordstretchfactor\fontdimen3\font minus
  \fontdimen4\font\relax}
\providecommand{\BIBforeignlanguage}[2]{{%
\expandafter\ifx\csname l@#1\endcsname\relax
\typeout{** WARNING: IEEEtran.bst: No hyphenation pattern has been}%
\typeout{** loaded for the language `#1'. Using the pattern for}%
\typeout{** the default language instead.}%
\else
\language=\csname l@#1\endcsname
\fi
#2}}
\providecommand{\BIBdecl}{\relax}
\BIBdecl

\bibitem{Metzger:2008ky}
R.~J. Metzger, O.~D. Klein, G.~R. Martin, and M.~A. Krasnow, ``{The branching
  programme of mouse lung development.}'' \emph{Nature}, vol. 453, no. 7196,
  pp. 745--750, Jun. 2008.

\bibitem{Blanc:2012ea}
P.~Blanc, K.~Coste, P.~Pouchin, J.-M. Aza{\"\i}s, L.~Blanchon, D.~Gallot, and
  V.~Sapin, ``{A role for mesenchyme dynamics in mouse lung branching
  morphogenesis.}'' \emph{PLoS ONE}, vol.~7, no.~7, p. e41643, 2012.

\bibitem{Watanabe:2004kr}
T.~Watanabe and F.~Costantini, ``{Real-time analysis of ureteric bud branching
  morphogenesis in vitro.}'' \emph{Developmental Biology}, vol. 271, no.~1, pp.
  98--108, Jul. 2004.

\bibitem{Meyer:2004de}
T.~N. Meyer, C.~Schwesinger, K.~T. Bush, R.~O. Stuart, D.~W. Rose, M.~M. Shah,
  D.~A. Vaughn, D.~L. Steer, and S.~K. Nigam, ``{Spatiotemporal regulation of
  morphogenetic molecules during in vitro branching of the isolated ureteric
  bud: toward a model of branching through budding in the developing kidney.}''
  \emph{Developmental Biology}, vol. 275, no.~1, pp. 44--67, Nov. 2004.

\bibitem{Costantini:2010p43730}
F.~Costantini and R.~Kopan, ``{Patterning a complex organ: branching
  morphogenesis and nephron segmentation in kidney development.}'' \emph{Dev
  Cell}, vol.~18, no.~5, pp. 698--712, May 2010.

\bibitem{Affolter:2009p25219}
M.~Affolter, R.~Zeller, and E.~Caussinus, ``{Tissue remodelling through
  branching morphogenesis},'' \emph{Nat Rev Mol Cell Biol}, 2009.

\bibitem{OchoaEspinosa:2012ux}
A.~Ochoa-Espinosa and M.~Affolter, ``{Branching Morphogenesis: From Cells to
  Organs and Back},'' \emph{Cold Spring Harbor perspectives in biology}, Jul.
  2012.

\bibitem{Lu:2008p47757}
P.~Lu and Z.~Werb, ``{Patterning mechanisms of branched organs},''
  \emph{Science}, vol. 322, no. 5907, pp. 1506--1509, Dec. 2008.

\bibitem{Davies2005}
J.~Davies, \emph{{Branching Morphogenesis}}.\hskip 1em plus 0.5em minus
  0.4em\relax Springer, 2005.

\bibitem{Harrison2010}
L.~G. Harrison, \emph{{The Shaping of Life}}.\hskip 1em plus 0.5em minus
  0.4em\relax Cambridge University Presss, 2010.

\bibitem{Bellusci:1997vq}
S.~Bellusci, J.~Grindley, H.~Emoto, N.~Itoh, and B.~L. Hogan, ``{Fibroblast
  growth factor 10 (FGF10) and branching morphogenesis in the embryonic mouse
  lung.}'' \emph{Development (Cambridge, England)}, vol. 124, no.~23, pp.
  4867--4878, Dec. 1997.

\bibitem{Wilhelm:2006p45512}
D.~Wilhelm and P.~Koopman, ``{The makings of maleness: towards an integrated
  view of male sexual development},'' \emph{Nat Rev Genet}, vol.~7, no.~8, pp.
  620--631, Aug. 2006.

\bibitem{Makarenkova:2009ha}
H.~P. Makarenkova, M.~P. Hoffman, A.~Beenken, A.~V. Eliseenkova, R.~Meech,
  C.~Tsau, V.~N. Patel, R.~A. Lang, and M.~Mohammadi, ``{Differential
  interactions of FGFs with heparan sulfate control gradient formation and
  branching morphogenesis.}'' \emph{Science signaling}, vol.~2, no.~88, p.
  ra55, 2009.

\bibitem{Hsu:2010ep}
J.~C.-f. Hsu and K.~M. Yamada, ``{Salivary gland branching
  morphogenesis--recent progress and future opportunities.}''
  \emph{International journal of oral science}, vol.~2, no.~3, pp. 117--126,
  Sep. 2010.

\bibitem{Bhushan:2001ua}
A.~Bhushan, N.~Itoh, S.~Kato, J.~P. Thiery, P.~Czernichow, S.~Bellusci, and
  R.~Scharfmann, ``{Fgf10 is essential for maintaining the proliferative
  capacity of epithelial progenitor cells during early pancreatic
  organogenesis.}'' \emph{Development (Cambridge, England)}, vol. 128, no.~24,
  pp. 5109--5117, Dec. 2001.

\bibitem{Pulkkinen:2003uy}
M.-A. Pulkkinen, B.~Spencer-Dene, C.~Dickson, and T.~Otonkoski, ``{The IIIb
  isoform of fibroblast growth factor receptor 2 is required for proper growth
  and branching of pancreatic ductal epithelium but not for differentiation of
  exocrine or endocrine cells.}'' \emph{Mechanisms of development}, vol. 120,
  no.~2, pp. 167--175, Feb. 2003.

\bibitem{Treanor:1996dq}
J.~J. Treanor, L.~Goodman, F.~de~Sauvage, D.~M. Stone, K.~T. Poulsen, C.~D.
  Beck, C.~Gray, M.~P. Armanini, R.~A. Pollock, F.~Hefti, H.~S. Phillips,
  A.~Goddard, M.~W. Moore, A.~Buj-Bello, A.~M. Davies, N.~Asai, M.~Takahashi,
  R.~Vandlen, C.~E. Henderson, and A.~Rosenthal, ``{Characterization of a
  multicomponent receptor for GDNF.}'' \emph{Nature}, vol. 382, no. 6586, pp.
  80--83, Jul. 1996.

\bibitem{Pichel:1996en}
J.~G. Pichel, L.~Shen, H.~Z. Sheng, A.~C. Granholm, J.~Drago, A.~Grinberg,
  E.~J. Lee, S.~P. Huang, M.~Saarma, B.~J. Hoffer, H.~Sariola, and H.~Westphal,
  ``{Defects in enteric innervation and kidney development in mice lacking
  GDNF.}'' \emph{Nature}, vol. 382, no. 6586, pp. 73--76, Jul. 1996.

\bibitem{Sanchez:1996cy}
M.~P. S{\'a}nchez, I.~Silos-Santiago, J.~Fris{\'e}n, B.~He, S.~A. Lira, and
  M.~Barbacid, ``{Renal agenesis and the absence of enteric neurons in mice
  lacking GDNF.}'' \emph{Nature}, vol. 382, no. 6586, pp. 70--73, Jul. 1996.

\bibitem{Tang:1998vf}
M.~J. Tang, D.~Worley, M.~Sanicola, and G.~R. Dressler, ``{The RET-glial
  cell-derived neurotrophic factor (GDNF) pathway stimulates migration and
  chemoattraction of epithelial cells.}'' \emph{J Cell Biol}, vol. 142, no.~5,
  pp. 1337--1345, Sep. 1998.

\bibitem{Lu:2008iu}
P.~Lu, A.~J. Ewald, G.~R. Martin, and Z.~Werb, ``{Genetic mosaic analysis
  reveals FGF receptor 2 function in terminal end buds during mammary gland
  branching morphogenesis.}'' \emph{Developmental Biology}, vol. 321, no.~1,
  pp. 77--87, Sep. 2008.

\bibitem{Li:2010dv}
R.~Li and B.~Bowerman, ``{Symmetry breaking in biology.}'' \emph{Cold Spring
  Harbor perspectives in biology}, vol.~2, no.~3, p. a003475, Mar. 2010.

\bibitem{Hirashima:2009p43515}
T.~Hirashima and Y.~Iwasa, ``{Mechanisms for split localization of Fgf10
  expression in early lung development},'' \emph{Developmental Dynamics}, 2009.

\bibitem{Kitaoka:1999uv}
H.~Kitaoka, R.~Takaki, and B.~Suki, ``{A three-dimensional model of the human
  airway tree.}'' \emph{Journal of applied physiology (Bethesda, Md. : 1985)},
  vol.~87, no.~6, pp. 2207--2217, Dec. 1999.

\bibitem{West:1986ty}
B.~J. West, V.~Bhargava, and A.~L. Goldberger, ``{Beyond the principle of
  similitude: renormalization in the bronchial tree.}'' \emph{Journal of
  applied physiology (Bethesda, Md. : 1985)}, vol.~60, no.~3, pp. 1089--1097,
  Mar. 1986.

\bibitem{Nelson:1990tw}
T.~R. Nelson, B.~J. West, and A.~L. Goldberger, ``{The fractal lung: universal
  and species-related scaling patterns.}'' \emph{Experientia}, vol.~46, no.~3,
  pp. 251--254, Mar. 1990.

\bibitem{Panico:1995gc}
J.~Panico and P.~Sterling, ``{Retinal neurons and vessels are not fractal but
  space-filling.}'' \emph{The Journal of comparative neurology}, vol. 361,
  no.~3, pp. 479--490, Oct. 1995.

\bibitem{Mauroy:2004ea}
B.~Mauroy, M.~Filoche, E.~R. WEIBEL, and B.~Sapoval, ``{An optimal bronchial
  tree may be dangerous.}'' \emph{Nature}, vol. 427, no. 6975, pp. 633--636,
  Feb. 2004.

\bibitem{Wilson:1967wu}
T.~A. Wilson, ``{Design of the bronchial tree.}'' \emph{Nature}, vol. 213, no.
  5077, pp. 668--669, Feb. 1967.

\bibitem{Bleuer:1988uo}
B.~H. Bleuer, ``{Morphometry of the human pulmonary acinus},'' \emph{The
  Anatomical Record}, 1988.

\bibitem{Lubkin1995}
S.~Lubkin and J.~Murray, ``{A mechanism for early branching in lung
  morphogenesis.}'' \emph{J Math Biol}, vol.~34, pp. 77--94, 1995.

\bibitem{Jha2011}
\BIBentryALTinterwordspacing
B.~Jha, L.~Cueto-Felgueroso, and R.~Juanes, ``{Quantifying mixing in viscously
  unstable porous media flows},'' \emph{Physical Review E}, vol.~84, no.~6,
  Dec. 2011. [Online]. Available:
  \url{http://link.aps.org/doi/10.1103/PhysRevE.84.066312}
\BIBentrySTDinterwordspacing

\bibitem{Unbekandt2008}
M.~Unbekandt, P.-M. del Moral, F.~G. Sala, S.~Bellusci, D.~Warburton, and
  V.~Fleury, ``{Tracheal occlusion increases the rate of epithelial branching
  of embryonic mouse lung via the FGF10-FGFR2b-Sprouty2 pathway.}''
  \emph{Mechanisms of development}, vol. 125, pp. 314--324, 2008.

\bibitem{Lin:2001vf}
Y.~Lin, S.~Zhang, M.~Rehn, P.~It{\"a}ranta, J.~Tuukkanen, R.~Helj{\"a}svaara,
  H.~Peltoketo, T.~Pihlajaniemi, and S.~Vainio, ``{Induced repatterning of type
  XVIII collagen expression in ureter bud from kidney to lung type: association
  with sonic hedgehog and ectopic surfactant protein C.}'' \emph{Development
  (Cambridge, England)}, vol. 128, no.~9, pp. 1573--1585, May 2001.

\bibitem{Nogawa:1987us}
H.~Nogawa and Y.~Nakanishi, ``{Mechanical aspects of the mesenchymal influence
  on epithelial branching morphogenesis of mouse salivary gland},''
  \emph{Development (Cambridge, England)}, 1987.

\bibitem{Mather1985}
\BIBentryALTinterwordspacing
J.~V. Maher, ``Development of viscous fingering patterns,'' \emph{Phys. Rev.
  Lett.}, vol.~54, pp. 1498--1501, Apr 1985. [Online]. Available:
  \url{http://link.aps.org/doi/10.1103/PhysRevLett.54.1498}
\BIBentrySTDinterwordspacing

\bibitem{Wan2008}
X.~Wan, Z.~Li, and S.~R. Lubkin, ``{Mechanics of mesenchymal contribution to
  clefting force in branching morphogenesis.}'' \emph{Biomechanics And Modeling
  In Mechanobiology}, vol.~7, pp. 417--426, 2008.

\bibitem{Lubkin:2008iy}
S.~R. Lubkin, ``{Branched organs: mechanics of morphogenesis by multiple
  mechanisms.}'' \emph{Current topics in developmental biology}, vol.~81, pp.
  249--268, 2008.

\bibitem{Lubkin2002}
S.~Lubkin and Z.~Li, ``{Force and deformation on branching rudiments: cleaving
  between hypotheses.}'' \emph{Biomech Model Mechanobiol}, vol.~1, pp. 5--16,
  2002.

\bibitem{Nogawa:1995wv}
H.~Nogawa and T.~Ito, ``{Branching morphogenesis of embryonic mouse lung
  epithelium in mesenchyme-free culture.}'' \emph{Development (Cambridge,
  England)}, vol. 121, no.~4, pp. 1015--1022, Apr. 1995.

\bibitem{Warburton:2010p46351}
D.~Warburton, A.~El-Hashash, G.~Carraro, C.~Tiozzo, F.~Sala, O.~Rogers,
  S.~De~Langhe, P.~J. Kemp, D.~Riccardi, J.~Torday, S.~Bellusci, W.~Shi, S.~R.
  Lubkin, and E.~Jesudason, ``{Lung organogenesis},'' \emph{Current topics in
  developmental biology}, vol.~90, pp. 73--158, 2010.

\bibitem{Moore2005}
K.~A. Moore, T.~Polte, S.~Huang, B.~Shi, E.~Alsberg, M.~E. Sunday, and D.~E.
  Ingber, ``{Control of basement membrane remodeling and epithelial branching
  morphogenesis in embryonic lung by Rho and cytoskeletal tension},''
  \emph{Developmental dynamics : an official publication of the American
  Association of Anatomists}, vol. 232, pp. 268--281, 2005.

\bibitem{Moore2002}
T.~M. Moore, W.~B. Shirah, P.~L. Khimenko, P.~Paisley, R.~N. Lausch, and A.~E.
  Taylor, ``{Involvement of CD40-CD40L signaling in postischemic lung
  injury},'' \emph{American journal of physiology Lung cellular and molecular
  physiology}, vol. 283, pp. L1255--62, 2002.

\bibitem{Kim:2012ih}
H.~Y. Kim and C.~M. Nelson, ``{Extracellular matrix and cytoskeletal dynamics
  during branching morphogenesis.}'' \emph{Organogenesis}, vol.~8, no.~2, pp.
  56--64, Apr. 2012.

\bibitem{Hartmann2006}
D.~Hartmann and T.~Miura, ``{Modelling in vitro lung branching morphogenesis
  during development.}'' \emph{Journal of theoretical biology}, vol. 242, pp.
  862--872, 2006.

\bibitem{Clement:2012fj}
R.~Cl{\'e}ment, P.~Blanc, B.~Mauroy, V.~Sapin, and S.~Douady, ``{Shape
  self-regulation in early lung morphogenesis.}'' \emph{PLoS ONE}, vol.~7,
  no.~5, p. e36925, 2012.

\bibitem{Menshykau:2012kg}
D.~Menshykau, C.~Kraemer, and D.~Iber, ``{Branch Mode Selection during Early
  Lung Development.}'' \emph{Plos Computational Biology}, vol.~8, no.~2, p.
  e1002377, Feb. 2012.

\bibitem{Min:1998p48579}
H.~Min, D.~M. Danilenko, S.~A. Scully, B.~Bolon, B.~D. Ring, J.~E. Tarpley,
  M.~DeRose, and W.~S. Simonet, ``{Fgf-10 is required for both limb and lung
  development and exhibits striking functional similarity to Drosophila
  branchless},'' \emph{Genes Dev}, vol.~12, no.~20, pp. 3156--3161, Oct. 1998.

\bibitem{Sekine:1999gd}
K.~Sekine, H.~Ohuchi, M.~Fujiwara, M.~Yamasaki, T.~Yoshizawa, T.~Sato,
  N.~Yagishita, D.~Matsui, Y.~Koga, N.~Itoh, and S.~Kato, ``{Fgf10 is essential
  for limb and lung formation.}'' \emph{Nat Genet}, vol.~21, no.~1, pp.
  138--141, Jan. 1999.

\bibitem{Pepicelli:1998p48620}
C.~V. Pepicelli, P.~M. Lewis, and A.~P. McMahon, ``{Sonic hedgehog regulates
  branching morphogenesis in the mammalian lung.}'' \emph{Current biology :
  CB}, vol.~8, no.~19, pp. 1083--1086, Sep. 1998.

\bibitem{Hartmann2007}
D.~Hartmann and T.~Miura, ``{Mathematical analysis of a free-boundary model for
  lung branching morphogenesis.}'' \emph{Mathematical medicine and biology : a
  journal of the IMA}, vol.~24, pp. 209--224, 2007.

\bibitem{Miura2002}
\BIBentryALTinterwordspacing
T.~Miura and K.~Shiota, ``{Depletion of FGF acts as a lateral inhibitory factor
  in lung branching morphogenesis in vitro.}'' \emph{Mechanisms of
  development}, vol. 116, no. 1-2, pp. 29--38, Aug. 2002. [Online]. Available:
  \url{http://www.ncbi.nlm.nih.gov/pubmed/12128203}
\BIBentrySTDinterwordspacing

\bibitem{Miura2009}
T.~Miura, D.~Hartmann, M.~Kinboshi, M.~Komada, M.~Ishibashi, and K.~Shiota,
  ``{The cyst-branch difference in developing chick lung results from a
  different morphogen diffusion coefficient},'' \emph{Mechanisms of
  development}, vol. 126, no. 3-4, pp. 160--172, Jan. 2009.

\bibitem{Clement:2012hw}
R.~Cl{\'e}ment, S.~Douady, and B.~Mauroy, ``{Branching geometry induced by lung
  self-regulated growth.}'' \emph{Physical biology}, vol.~9, no.~6, p. 066006,
  Nov. 2012.

\bibitem{Mailleux:2001vf}
A.~A. Mailleux, D.~Tefft, D.~Ndiaye, N.~Itoh, J.~P. Thiery, D.~Warburton, and
  S.~Bellusci, ``{Evidence that SPROUTY2 functions as an inhibitor of mouse
  embryonic lung growth and morphogenesis.}'' \emph{Mechanisms of development},
  vol. 102, no. 1-2, pp. 81--94, Apr. 2001.

\bibitem{Bragg:2001ws}
A.~D. Bragg, H.~L. Moses, and R.~Serra, ``{Signaling to the epithelium is not
  sufficient to mediate all of the effects of transforming growth factor beta
  and bone morphogenetic protein 4 on murine embryonic lung development.}''
  \emph{Mechanisms of development}, vol. 109, no.~1, pp. 13--26, Nov. 2001.

\bibitem{Gleghorn:2012el}
J.~P. Gleghorn, J.~Kwak, A.~L. Pavlovich, and C.~M. Nelson, ``{Inhibitory
  morphogens and monopodial branching of the embryonic chicken lung.}''
  \emph{Developmental dynamics : an official publication of the American
  Association of Anatomists}, Mar. 2012.

\bibitem{Nelson:2006gn}
C.~M. Nelson, M.~M. Vanduijn, J.~L. Inman, D.~A. Fletcher, and M.~J. Bissell,
  ``{Tissue geometry determines sites of mammary branching morphogenesis in
  organotypic cultures.}'' \emph{Science}, vol. 314, no. 5797, pp. 298--300,
  Oct. 2006.

\bibitem{Celliere2012}
\BIBentryALTinterwordspacing
G.~Celliere, D.~Menshykau, and D.~Iber, ``{Simulations demonstrate a simple
  network to be sufficient to control branch point selection, smooth muscle and
  vasculature formation during lung branching morphogenesis},'' \emph{Biology
  Open}, Jul. 2012. [Online]. Available:
  \url{http://bio.biologists.org/cgi/doi/10.1242/bio.20121339}
\BIBentrySTDinterwordspacing

\bibitem{MurrayBook}
J.~D. Murray, \emph{{Mathematical Biology. 3rd edition in 2 volumes:
  Mathematical Biology: II. Spatial Models and Biomedical Applications.}}\hskip
  1em plus 0.5em minus 0.4em\relax Springer, 2003.

\bibitem{Lazarus:2011cp}
A.~Lazarus, P.-M. del Moral, O.~Ilovich, E.~Mishani, D.~Warburton, and
  E.~Keshet, ``{A perfusion-independent role of blood vessels in determining
  branching stereotypy of lung airways.}'' \emph{Development (Cambridge,
  England)}, vol. 138, no.~11, pp. 2359--2368, Jun. 2011.

\bibitem{Tang:2011ii}
N.~Tang, W.~F. Marshall, M.~McMahon, R.~J. Metzger, and G.~R. Martin,
  ``{Control of mitotic spindle angle by the RAS-regulated ERK1/2 pathway
  determines lung tube shape.}'' \emph{Science}, vol. 333, no. 6040, pp.
  342--345, Jul. 2011.

\bibitem{Majumdar:2003tp}
A.~Majumdar, S.~Vainio, A.~Kispert, J.~McMahon, and A.~P. McMahon, ``{Wnt11 and
  Ret/Gdnf pathways cooperate in regulating ureteric branching during
  metanephric kidney development.}'' \emph{Development (Cambridge, England)},
  vol. 130, no.~14, pp. 3175--3185, Jul. 2003.

\bibitem{SimsLucas:2009bq}
S.~Sims-Lucas, C.~Argyropoulos, K.~Kish, K.~McHugh, J.~F. Bertram, R.~Quigley,
  and C.~M. Bates, ``{Three-dimensional imaging reveals ureteric and
  mesenchymal defects in Fgfr2-mutant kidneys.}'' \emph{Journal of the American
  Society of Nephrology : JASN}, vol.~20, no.~12, pp. 2525--2533, Dec. 2009.

\bibitem{Pepicelli1997}
C.~Pepicelli, A.~Kispert, D.~Rowitch, and A.~McMahon, ``{GDNF induces branching
  and increased cell proliferation in the ureter of the mouse.}'' \emph{Dev
  Biol}, vol. 192, no.~1, pp. 193--198, 1997.

\bibitem{Schuchardt:1994hg}
A.~Schuchardt, V.~D'Agati, L.~Larsson-Blomberg, F.~Costantini, and V.~Pachnis,
  ``{Defects in the kidney and enteric nervous system of mice lacking the
  tyrosine kinase receptor Ret.}'' \emph{Nature}, vol. 367, no. 6461, pp.
  380--383, Jan. 1994.

\bibitem{Michos:2010p43732}
O.~Michos, C.~Cebrian, D.~Hyink, U.~Grieshammer, L.~Williams, V.~D'Agati, J.~D.
  Licht, G.~R. Martin, and F.~Costantini, ``{Kidney Development in the Absence
  of Gdnf and Spry1 Requires Fgf10},'' \emph{PLoS genetics}, vol.~6, no.~1, p.
  e1000809, Jan. 2010.

\bibitem{Lu2009}
B.~C. Lu, C.~Cebrian, X.~Chi, S.~Kuure, R.~Kuo, C.~M. Bates, S.~Arber,
  J.~Hassell, L.~MacNeil, M.~Hoshi, S.~Jain, N.~Asai, M.~Takahashi, K.~M.
  Schmidt-Ott, J.~Barasch, V.~D'Agati, and F.~Costantini, ``{Etv4 and Etv5 are
  required downstream of GDNF and Ret for kidney branching morphogenesis.}''
  \emph{Nature genetics}, vol.~41, no.~12, pp. 1295--302, 2009.

\bibitem{Tang:2002it}
M.-J. Tang, Y.~Cai, S.-J. Tsai, Y.-K. Wang, and G.~R. Dressler, ``{Ureteric bud
  outgrowth in response to RET activation is mediated by phosphatidylinositol
  3-kinase.}'' \emph{Developmental Biology}, vol. 243, no.~1, pp. 128--136,
  Mar. 2002.

\bibitem{Sariola:2003jo}
H.~Sariola and M.~Saarma, ``{Novel functions and signalling pathways for
  GDNF.}'' \emph{Journal of cell science}, vol. 116, no. Pt 19, pp. 3855--3862,
  Oct. 2003.

\bibitem{Hirashima:2009er}
T.~Hirashima, Y.~Iwasa, and Y.~Morishita, ``{Dynamic modeling of branching
  morphogenesis of ureteric bud in early kidney development.}'' \emph{Journal
  of theoretical biology}, vol. 259, no.~1, pp. 58--66, Jul. 2009.

\bibitem{Menshykau:nMxfL07C}
D.~Menshykau and D.~Iber, ``{Kidney branching morphogenesis under the control
  of a ligand-receptor based Turing mechanism},'' \emph{Physical biology}.

\bibitem{Iber:2013vf}
D.~Iber, S.~Tanaka, P.~Fried, P.~Germann, and D.~Menshykau, ``{Simulating
  Tissue Morphogenesis and Signaling },'' in \emph{Tissue Morphogenesis:
  Methods and Protocols}, C.~M. Nelson, Ed.\hskip 1em plus 0.5em minus
  0.4em\relax Methods in Molecular Biology (Springer), Dec. 2013.

\bibitem{Blanc2012}
\BIBentryALTinterwordspacing
P.~Blanc, K.~Coste, P.~Pouchin, J.-M. Aza\"{\i}s, L.~Blanchon, D.~Gallot, and
  V.~Sapin, ``{A role for mesenchyme dynamics in mouse lung branching
  morphogenesis.}'' \emph{PloS one}, vol.~7, no.~7, p. e41643, Jan. 2012.
  [Online]. Available: \url{http://www.ncbi.nlm.nih.gov/pubmed/22844507}
\BIBentrySTDinterwordspacing

\bibitem{Kondo:2010bx}
S.~Kondo and T.~Miura, ``{Reaction-diffusion model as a framework for
  understanding biological pattern formation.}'' \emph{Science}, vol. 329, no.
  5999, pp. 1616--1620, Sep. 2010.

\bibitem{Hoefer:tr}
T.~Hoefer and P.~Maini, ``{Turing patterns in fish skin?}'' \emph{Nature}, vol.
  380, p. 678, 1996.

\bibitem{Akam:1989kk}
M.~Akam, ``{Drosophila development: making stripes inelegantly.}''
  \emph{Nature}, vol. 341, no. 6240, pp. 282--283, Sep. 1989.

\bibitem{Mailleux2005}
A.~A. Mailleux, ``Fgf10 expression identifies parabronchial smooth muscle cell
  progenitors and is required for their entry into the smooth muscle cell
  lineage,'' \emph{Development}, vol. 132, no.~9, pp. 2157--2166, 2005.

\bibitem{Ramasamy2007}
S.~K. Ramasamy, A.~A. Mailleux, V.~V. Gupte, F.~Mata, F.~G. Sala, J.~M.
  Veltmaat, P.~M. Del~Moral, S.~De~Langhe, S.~Parsa, L.~K. Kelly, R.~Kelly,
  W.~Shia, E.~Keshet, P.~Minoo, D.~Warburton, and S.~Bellusci, ``{Fgf10 dosage
  is critical for the amplification of epithelial cell progenitors and for the
  formation of multiple mesenchymal lineages during lung development.}''
  \emph{Dev Biol}, vol. 307, pp. 237--247, 2007.

\bibitem{Majumdar2003}
A.~Majumdar, S.~Vainio, A.~Kispert, J.~McMahon, and A.~P. McMahon, ``{Wnt11 and
  Ret/Gdnf pathways cooperate in regulating ureteric branching during
  metanephric kidney development.}'' \emph{Development (Cambridge, England)},
  vol. 130, no.~14, pp. 3175--3185, Jul. 2003.

\bibitem{Larsen:2010ki}
M.~Larsen, K.~M. Yamada, and K.~Musselmann, ``{Systems analysis of salivary
  gland development and disease.}'' \emph{Wiley interdisciplinary reviews.
  Systems biology and medicine}, vol.~2, no.~6, pp. 670--682, Nov. 2010.

\bibitem{Haara:2011kn}
O.~H{\"a}{\"a}r{\"a}, S.~Fujimori, R.~Schmidt-Ullrich, C.~Hartmann,
  I.~Thesleff, and M.~L. Mikkola, ``{Ectodysplasin and Wnt pathways are
  required for salivary gland branching morphogenesis.}'' \emph{Development
  (Cambridge, England)}, vol. 138, no.~13, pp. 2681--2691, Jul. 2011.

\bibitem{Jaskoll:2004eo}
T.~Jaskoll, T.~Leo, D.~Witcher, M.~Ormestad, J.~Astorga, P.~Bringas,
  P.~Carlsson, and M.~Melnick, ``{Sonic hedgehog signaling plays an essential
  role during embryonic salivary gland epithelial branching morphogenesis.}''
  \emph{Developmental dynamics : an official publication of the American
  Association of Anatomists}, vol. 229, no.~4, pp. 722--732, Apr. 2004.

\bibitem{Jaskoll:2004hl}
T.~Jaskoll, D.~Witcher, L.~Toreno, P.~Bringas, A.~M. Moon, and M.~Melnick,
  ``{FGF8 dose-dependent regulation of embryonic submandibular salivary gland
  morphogenesis.}'' \emph{Developmental Biology}, vol. 268, no.~2, pp.
  457--469, Apr. 2004.

\bibitem{Patel:2006ed}
V.~N. Patel, I.~T. Rebustini, and M.~P. Hoffman, ``{Salivary gland branching
  morphogenesis.}'' \emph{Differentiation; research in biological diversity},
  vol.~74, no.~7, pp. 349--364, Sep. 2006.

\bibitem{Villasenor:2010iz}
A.~Villasenor, D.~C. Chong, M.~Henkemeyer, and O.~Cleaver, ``{Epithelial
  dynamics of pancreatic branching morphogenesis.}'' \emph{Development
  (Cambridge, England)}, vol. 137, no.~24, pp. 4295--4305, Dec. 2010.

\bibitem{Benitez:2012hc}
C.~M. Benitez, W.~R. Goodyer, and S.~K. Kim, ``{Deconstructing pancreas
  developmental biology.}'' \emph{Cold Spring Harbor perspectives in biology},
  vol.~4, no.~6, Jun. 2012.

\bibitem{Puri:2007jp}
S.~Puri and M.~Hebrok, ``{Dynamics of embryonic pancreas development using
  real-time imaging.}'' \emph{Developmental Biology}, vol. 306, no.~1, pp.
  82--93, Jun. 2007.

\bibitem{Kawahira:2003p43684}
H.~Kawahira, N.~H. Ma, E.~S. Tzanakakis, A.~P. McMahon, P.-T. Chuang, and
  M.~Hebrok, ``{Combined activities of hedgehog signaling inhibitors regulate
  pancreas development.}'' \emph{Development (Cambridge, England)}, vol. 130,
  no.~20, pp. 4871--4879, Oct. 2003.

\bibitem{Setty:2008ea}
Y.~Setty, I.~R. Cohen, Y.~Dor, and D.~Harel, ``{Four-dimensional realistic
  modeling of pancreatic organogenesis.}'' \emph{PNAS}, vol. 105, no.~51, pp.
  20\,374--20\,379, Dec. 2008.

\bibitem{Petiot:2005dr}
A.~Petiot, C.~L. Perriton, C.~Dickson, and M.~J. Cohn, ``{Development of the
  mammalian urethra is controlled by Fgfr2-IIIb.}'' \emph{Development
  (Cambridge, England)}, vol. 132, no.~10, pp. 2441--2450, May 2005.

\bibitem{Donjacour:2003ty}
A.~A. Donjacour, A.~A. Thomson, and G.~R. Cunha, ``{FGF-10 plays an essential
  role in the growth of the fetal prostate.}'' \emph{Developmental Biology},
  vol. 261, no.~1, pp. 39--54, Sep. 2003.

\bibitem{Macias:2012iv}
H.~Macias and L.~Hinck, ``{Mammary Gland Development.}'' \emph{Wiley
  interdisciplinary reviews. Developmental biology}, vol.~1, no.~4, pp.
  533--557, Jul. 2012.

\bibitem{Kratochwil:1976tq}
K.~Kratochwil and P.~Schwartz, ``{Tissue interaction in androgen response of
  embryonic mammary rudiment of mouse: identification of target tissue for
  testosterone.}'' \emph{Proceedings of the National Academy of Sciences of the
  United States of America}, vol.~73, no.~11, pp. 4041--4044, Nov. 1976.

\bibitem{Wysolmerski:1998vu}
J.~J. Wysolmerski, W.~M. Philbrick, M.~E. Dunbar, B.~Lanske, H.~Kronenberg, and
  A.~E. Broadus, ``{Rescue of the parathyroid hormone-related protein knockout
  mouse demonstrates that parathyroid hormone-related protein is essential for
  mammary gland development.}'' \emph{Development (Cambridge, England)}, vol.
  125, no.~7, pp. 1285--1294, Apr. 1998.

\bibitem{Hens:2007cv}
J.~R. Hens, P.~Dann, J.-P. Zhang, S.~Harris, G.~W. Robinson, and
  J.~Wysolmerski, ``{BMP4 and PTHrP interact to stimulate ductal outgrowth
  during embryonic mammary development and to inhibit hair follicle
  induction.}'' \emph{Development (Cambridge, England)}, vol. 134, no.~6, pp.
  1221--1230, Mar. 2007.

\bibitem{Gjorevski:2011hg}
N.~Gjorevski and C.~M. Nelson, ``{Integrated morphodynamic signalling of the
  mammary gland.}'' \emph{Nat Rev Mol Cell Biol}, vol.~12, no.~9, pp. 581--593,
  Sep. 2011.

\bibitem{Park:1998fz}
W.~Y. Park, B.~Miranda, D.~Lebeche, G.~Hashimoto, and W.~V. Cardoso, ``{FGF-10
  is a chemotactic factor for distal epithelial buds during lung
  development.}'' \emph{Developmental Biology}, vol. 201, no.~2, pp. 125--134,
  Sep. 1998.

\bibitem{Pierce:1993va}
D.~F. Pierce, M.~D. Johnson, Y.~Matsui, S.~D. Robinson, L.~I. Gold, A.~F.
  Purchio, C.~W. Daniel, B.~L. Hogan, and H.~L. Moses, ``{Inhibition of mammary
  duct development but not alveolar outgrowth during pregnancy in transgenic
  mice expressing active TGF-beta 1.}'' \emph{Genes Dev}, vol.~7, no. 12A, pp.
  2308--2317, Dec. 1993.

\bibitem{Macias:2011jt}
H.~Macias, A.~Moran, Y.~Samara, M.~Moreno, J.~E. Compton, G.~Harburg,
  P.~Strickland, and L.~Hinck, ``{SLIT/ROBO1 signaling suppresses mammary
  branching morphogenesis by limiting basal cell number.}'' \emph{Dev Cell},
  vol.~20, no.~6, pp. 827--840, Jun. 2011.

\bibitem{Roarty:2007hq}
K.~Roarty and R.~Serra, ``{Wnt5a is required for proper mammary gland
  development and TGF-beta-mediated inhibition of ductal growth.}''
  \emph{Development (Cambridge, England)}, vol. 134, no.~21, pp. 3929--3939,
  Nov. 2007.

\bibitem{Kratochwil:1969tj}
K.~Kratochwil, ``{Organ specificity in mesenchymal induction demonstrated in
  the embryonic development of the mammary gland of the mouse.}''
  \emph{Developmental Biology}, vol.~20, no.~1, pp. 46--71, Jul. 1969.

\bibitem{Pozzi:2011eu}
A.~Pozzi and R.~Zent, ``{Extracellular matrix receptors in branched organs.}''
  \emph{Curr Opin Cell Biol}, vol.~23, no.~5, pp. 547--553, Oct. 2011.

\bibitem{Nelson:2012ku}
C.~M. Nelson, ``{Symmetry breaking during morphogenesis in the embryo and in
  engineered tissues},'' \emph{AIChE Journal}, vol.~58, no.~12, pp. 3608--3613,
  Oct. 2012.

\bibitem{Gjorevski:2010kr}
N.~Gjorevski and C.~M. Nelson, ``{The mechanics of development: Models and
  methods for tissue morphogenesis.}'' \emph{Birth defects research Part C,
  Embryo today : reviews}, vol.~90, no.~3, pp. 193--202, Sep. 2010.

\bibitem{Choquet:2003fk}
D.~Choquet and A.~Triller, ``{The role of receptor diffusion in the
  organization of the postsynaptic membrane.}'' \emph{Nature reviews.
  Neuroscience}, vol.~4, no.~4, pp. 251--265, Apr. 2003.

\bibitem{Ries:2009p20248}
J.~Ries, S.~R. Yu, M.~Burkhardt, M.~Brand, and P.~Schwille, ``{Modular scanning
  FCS quantifies receptor-ligand interactions in living multicellular
  organisms},'' \emph{Nature methods}, vol.~6, no.~9, pp. 643--645, Sep. 2009.

\bibitem{Kumar2010}
M.~Kumar, M.~S. Mommer, and V.~Sourjik, ``{Mobility of Cytoplasmic, Membrane,
  and DNA-Binding Proteins in Escherichia coli.}'' \emph{Biophysical journal},
  vol.~98, no.~4, pp. 552--559, 2010.

\bibitem{Hebert2005}
B.~Hebert, S.~Costantino, and P.~Wiseman, ``{Spatiotemporal image correlation
  Spectroscopy (STICS) theory, verification, and application to protein
  velocity mapping in living CHO cells.}'' \emph{Biophysical journal}, vol.~88,
  no.~5, pp. 3601--3614, 2005.

\bibitem{Weaver2010}
M.~Weaver, L.~Batts, and B.~Hogan, ``Tissue interactions pattern the mesenchyme
  of the embryonic mouse lung.'' \emph{Dev Biol}, vol. 258, no.~1, pp.
  169--184, 2010.

\bibitem{Costantini2010}
F.~Costantini and R.~Kopan, ``{Patterning a complex organ: branching
  morphogenesis and nephron segmentation in kidney development},'' \emph{Dev
  Cell}, vol.~18, no.~5, pp. 698--712, May 2010.

\bibitem{Bansal1997}
R.~Bansal and S.~Pfeiffer, ``Regulation of oligodendrocyte differentiation by
  fibroblast growth factors.'' \emph{Adv Exp Med Biol}, vol. 429, pp. 69--77,
  1997.

\bibitem{Estival2010}
A.~Estival, V.~Monzat, K.~Miquel, F.~Gaubert, E.~Hollande, M.~Korc, N.~Vaysse,
  and F.~Clementem, ``Differential regulation of fibroblast growth factor (fgf)
  receptor-1 mrna and protein by two molecular forms of basic fgf. modulation
  of fgfr-1 mrna stability.'' \emph{Dev Cell}, vol.~18, no.~5, pp. 698--712,
  2010.

\bibitem{Ota2010}
S.~Ota, N.~Tonou-Fujimori, N.~Tonou-Fujimori, Y.~Nakayama, Y.~Ito, A.~Kawamura,
  and K.~Yamasu, ``Fgf receptor gene expression and its regulation by fgf
  signaling during early zebrafish development.'' \emph{Genesis}, vol.~48,
  no.~12, pp. 707--716, 2010.

\bibitem{Zakrzewska:2013gt}
M.~Zakrzewska, E.~M. Haugsten, B.~Nadratowska-Wesolowska, A.~Oppelt,
  B.~Hausott, Y.~Jin, J.~Otlewski, J.~Wesche, and A.~Wiedlocha, ``{ERK-Mediated
  Phosphorylation of Fibroblast Growth Factor Receptor 1 on Ser777 Inhibits
  Signaling.}'' \emph{Science signaling}, vol.~6, no. 262, p. ra11, 2013.

\bibitem{Merino:1998ha}
R.~Merino, Y.~Ga{\~n}an, D.~Macias, A.~N. Economides, K.~T. Sampath, and J.~M.
  Hurle, ``{Morphogenesis of digits in the avian limb is controlled by FGFs,
  TGFbetas, and noggin through BMP signaling.}'' \emph{Developmental Biology},
  vol. 200, no.~1, pp. 35--45, Aug. 1998.

\bibitem{Badugu:2012ho}
A.~Badugu, C.~Kraemer, P.~Germann, D.~Menshykau, and D.~Iber, ``{Digit
  patterning during limb development as a result of the BMP-receptor
  interaction.}'' \emph{Scientific reports}, vol.~2, p. 991, 2012.

\bibitem{Gierer:1972vq}
A.~Gierer and H.~Meinhardt, ``{A theory of biological pattern formation.}''
  \emph{Kybernetik}, vol.~12, no.~1, pp. 30--39, Dec. 1972.

\bibitem{Schnakenberg:1979td}
J.~Schnakenberg, ``{Simple chemical reaction systems with limit cycle
  behaviour.}'' \emph{Journal of theoretical biology}, vol.~81, no.~3, pp.
  389--400, Dec. 1979.

\bibitem{Meinhardt:1976vc}
H.~Meinhardt, ``{Morphogenesis of lines and nets.}'' \emph{Differentiation;
  research in biological diversity}, vol.~6, no.~2, pp. 117--123, Aug. 1976.

\bibitem{Sharpe2002}
J.~Sharpe, U.~Ahlgren, P.~Perry, B.~Hill, A.~Ross, R.~B. Jacob
  Hecksher-Sorensen, and D.~Davidson, ``{Optical Projection Tomography as a
  Tool for 3D Microscopy and Gene Expression Studies.}'' \emph{Science}, vol.
  296, pp. 541--545, 2002.

\bibitem{Verveer:2007ck}
P.~J. Verveer, J.~Swoger, F.~Pampaloni, K.~Greger, M.~Marcello, and E.~H.~K.
  Stelzer, ``{High-resolution three-dimensional imaging of large specimens with
  light sheet-based microscopy.}'' \emph{Nature methods}, vol.~4, no.~4, pp.
  311--313, Apr. 2007.

\bibitem{Truong:2011cn}
T.~V. Truong, W.~Supatto, D.~S. Koos, J.~M. Choi, and S.~E. Fraser, ``{Deep and
  fast live imaging with two-photon scanned light-sheet microscopy.}''
  \emph{Nature methods}, vol.~8, no.~9, pp. 757--760, Sep. 2011.

\bibitem{Menshykau:2012vg}
D.~Menshykau and D.~Iber, ``{Simulating Organogenesis with Comsol: Interacting
  and Deforming Domains},'' \emph{Proceedings of COMSOL Conference 2012}, Sep.
  2012.

\bibitem{Germann:bT_kMV7D}
P.~Germann, D.~Menshykau, S.~Tanaka, and D.~Iber, ``{Simulating organogensis in
  Comsol},'' in \emph{Proceedings of COMSOL Conference 2011}, 2011.

\bibitem{Kim:2013db}
H.~Y. Kim, V.~D. Varner, and C.~M. Nelson, ``{Apical constriction initiates new
  bud formation during monopodial branching of the embryonic chicken lung.}''
  \emph{Development (Cambridge, England)}, Jul. 2013.

\bibitem{Schnatwinkel:2013kia}
C.~Schnatwinkel and L.~Niswander, ``{Multiparametric image analysis of
  lung-branching morphogenesis.}'' \emph{Developmental dynamics : an official
  publication of the American Association of Anatomists}, vol. 242, no.~6, pp.
  622--637, Jun. 2013.
  
  \bibitem{Hsu:2013hu}
J.~C Hsu, H. Koo, J.~S. Harunaga, K. Matsumoto, A.~D. Doyle, and K.~M. Yama, ``{Region-specific epithelial cell dynamics during branching morphogenesis.}''
  \emph{Dev Dyn}, Jun. 2013.




\end{thebibliography}


\end{document}